\documentclass[twocolumn,tighten,usenatbib,useAMS,fleqn,usenames,dvipsnames]{aastex63}

\usepackage{amsmath}
\usepackage{cancel}
\usepackage{nccmath}

\definecolor{enth_color}{HTML}{293132}
\definecolor{cond_color}{HTML}{702a70}
\definecolor{cool_color}{HTML}{1489b2}
\definecolor{xvis_color}{HTML}{ef476f}
\definecolor{zvis_color}{HTML}{ff9c44}
\definecolor{work_color}{HTML}{87b04b}

\definecolor{internationalkleinblue}{rgb}{0.0, 0.4, 0.8}
\definecolor{britishracinggreen}{rgb}{0.0, 0.5, 0.3}
\hypersetup{colorlinks=true, citecolor=internationalkleinblue, linkcolor=britishracinggreen, urlcolor=internationalkleinblue}

\graphicspath{{./}{figures/}}
\received{2022 November 2}
\revised{2023 February 24}
\accepted{2023 March 22}
\published{2023 June 22}

\newcommand{\tcool}{t_{\rm cool}}
\newcommand{\vrel}{v_{\rm rel}}
\newcommand{\Mach}{\mathcal{M}}
\newcommand{\Machrel}{\mathcal{M}_{\rm rel}}

\newcommand{\vz}{v_z}
\newcommand{\vx}{v_x}

\newcommand{\dt}[1]{\frac{\partial #1}{\partial t}}

\newcommand{\lr}[1]{\left( #1 \right)}

\newcommand{\dz}[1]{\frac{d #1}{d z}}

\newcommand{\dT}{\frac{d T}{d z}}
\newcommand{\dvz}{\frac{d v_z}{d z}}
\newcommand{\dvx}{\frac{d v_x}{d z}}
\newcommand{\ddT}{\frac{d^2 T}{d z^2}}
\newcommand{\ddvz}{\frac{d^2 v_z}{d z^2}}
\newcommand{\ddvx}{\frac{d^2 v_x}{d z^2}}
\newcommand{\dk}{\frac{d \kappa}{d z}}

\newcommand{\del}{\boldsymbol\nabla}
\newcommand{\Tv}{\boldsymbol \Pi}
\newcommand{\Qc}{\boldsymbol {\sf Q}}
\newcommand{\vv}{\boldsymbol v}
\newcommand{\PR}{\operatorname{\mathit{P\kern-.1em r}}}
\newcommand{\nuc}{\nu_c}
\newcommand{\tsh}{t_{\rm sh}}
\newcommand{\Tpeak}{T_{\rm peak}}
\newcommand{\betalo}{\beta_{\rm cold}}
\newcommand{\betahi}{\beta_{\rm hot}}
\newcommand{\Edotcool}{\dot{E}_{\rm cool}}
\newcommand{\Qcool}{Q_{\rm cool}}
\newcommand{\Hvisc}{\mathcal{H}_{\rm visc}}

\defcitealias{Fielding:2020}{FOBJ20}
\defcitealias{Gronke:2018}{GO18}
\defcitealias{Gronke:2020}{GO20}
\defcitealias{Tan:2021}{TOG21}

\shorttitle{Analytic Model for TRML}
\shortauthors{Chen, Fielding, \& Bryan}

\begin{document}

\title{The Anatomy of a Turbulent Radiative Mixing Layer: Insights from an Analytic Model with Turbulent Conduction and Viscosity}

\correspondingauthor{Zirui Chen}
\email{ziruichen@ucsb.edu}

\author[0000-0001-8755-3836]{Zirui Chen}
\affiliation{Department of Astronomy, Columbia University, 550 W 120th Street, New York, NY 10027, USA}
\affiliation{Department of Physics, University of California, Santa Barbara, Santa Barbara, CA 93106, USA}

\author[0000-0003-3806-8548]{Drummond B. Fielding}
\affiliation{Center for Computational Astrophysics, Flatiron Institute, 162 5th Ave, New York, NY 10010, USA}

\author[0000-0003-2630-9228]{Greg L. Bryan}
\affiliation{Department of Astronomy, Columbia University, 550 W 120th Street, New York, NY 10027, USA}
\affiliation{Center for Computational Astrophysics, Flatiron Institute, 162 5th Ave, New York, NY 10010, USA}

\begin{abstract}
Turbulent Radiative Mixing Layers (TRMLs) form at the interface of cold, dense gas and hot, diffuse gas in motion with each other. TRMLs are ubiquitous in and around galaxies on a variety of scales, including galactic winds and the circumgalactic medium. They host the intermediate temperature gases that are efficient in radiative cooling, thus play a crucial role in controlling the cold gas supply, phase structure, and spectral features of galaxies. In this work, we develop an intuitive analytic 1.5 dimensional model for TRMLs that includes a simple parameterization of the effective turbulent conductivity and viscosity and a piece-wise power-law cooling curve. Our analytic model reproduces the mass flux, total cooling, and phase structure of 3D simulations of TRMLs at a fraction of the computational cost. It also reveals essential insights into the physics of TRMLs, particularly the importance of the viscous dissipation of relative kinetic energy in balancing radiative cooling as the shear Mach number approaches unity.  This dissipation takes place both in the intermediate temperature phase, which reduces the enthalpy flux from the hot phase, and in the cold phase, which enhances radiative cooling. Additionally, our model provides a fast and easy way of computing the column density and surface brightness of TRMLs, which can be directly linked to observations.
\end{abstract}


\section{Introduction} \label{sec:intro}

The interaction and mixing of a hot wind and cold gas clouds is a ubiquitous situation in many astrophysical systems. In particular, both simulations \citep[e.g.][]{Kim:2018,Schneider:2020} and observations \citep[e.g.][]{StricklandHeckman:2009,Rupke:2018} suggest that astrophysical gases in and around galaxies are highly multiphase. That is, cold and hot gases coexist on a wide range of scales. These cold and hot phases are often in pressure equilibrium and move relative to each other. At the interface of the two phases Kelvin-Helmholtz instability (KHI) drives turbulent mixing which populates the intermediate temperatures that radiate efficiently, thus forming turbulent radiative mixing layers (TRMLs). TRMLs are essential in controlling the phase structure and evolution of galaxies on a variety of scales, including the interstellar medium \citep{Kim:2013, Audit:2010} and galactic winds \citep{Gronke:2020, Fielding:2022} on the parsec scale, supernova remnants and superbubbles \citep{Kim:2017, Fielding:2018, Orr:2022} on the kilo-parsec scale, and the circumgalactic medium \citep{Fielding:2017a}, the intracluster medium, and cosmic filaments \citep{Mandelker:2020b} on the mega-parsec scale. Furthermore, the theory governing TRMLs and the evolution of cold clouds in hot winds has close parallels in many other subfields of physics, including turbulent combustion in engines \citep{Tan:2021}, stellar interiors \citep{Meakin:2007}, solar corona \citep{Hillier:2019}, stellar wind bubbles \citep{Lancaster:2021}, supernovae \citep{Niemeyer:1997}, and interactions between clouds and their host planetary atmospheres \citep{Pauluis:2011}. 

Recently, TRMLs have been established as a crucial component of galaxy evolution. \cite{Gronke:2018} included radiative cooling in their 3D hydrodynamical cloud-crushing simulations and argued that TRMLs provide a viable solution to the cloud entrainment problem, in which simple arguments neglecting cooling predict that cold clouds should be shredded apart prior to being accelerated to high velocities that they are observed to have in galactic winds and the circumgalactic medium \citep[e.g.,][]{Zhang:2017}. Intermediate temperature gases in TRMLs consist of partially ionized atoms that are most conducive to radiative cooling, which means they convert hot gas to cold gas rapidly, allowing the original cold cloud to survive or even grow when it gets entrained in the hot wind. Later works confirmed the conclusion of cloud growth in the presence of radiative cooling when $\left. t_{\rm cool, mix} \right/ t_{\rm cc} \lesssim 1$, where $t_{\rm cool, mix}$ is the characteristic cooling time of the mixed gas\footnote{There is much debate in the literature about what is the appropriate cooling time to use. Here we adopt the the minimum cooling time at the pressure of the system. Most other choices introduce a qualitative, as opposed to quantitative change.}, and $t_{\rm cc}$ is the cloud crushing time (e.g., \citealt{Kanjilal:2021, Abruzzo:2021, Gronke:2021}, see \citealt{Li:2020, Sparre:2020} for a somewhat different interpretation predicated on the central importance of the hot phase cooling time).

Thus, TRMLs dictate the fate of cold clouds and play a crucial role in regulating the amount of cold gas that is available for star formation and black hole growth in galaxies. Turbulent Radiative Mixing Layers are also essential to observations because they host the intermediate temperature gases that contribute a significant amount of the spectral features (see e.g. \citealt{Gronke:2020} on how a significant amount of emission from a cloud comes from TRMLs) commonly seen in both emission \citep[e.g.,][]{Hayes:2016,ReichardtChu:2022} and absorption \citep[e.g.,][]{Werk:2014,Qu:2022}.

There has been significant work focusing on the interface to study TRMLs specifically. \cite{Begelman:1990} were amongst the earliest, and constructed a model where the advected enthalpy from the hot phase is balanced by cooling and discussed the rich emission and absorption features that are produced by the mixing layers. More recently, Kwak and collaborators \citep{Kwak:2010, Kwak:2011, Henley:2012, Kwak:2015} performed 2D simulations of TRMLs. They compared their results with observables, including line ratios and column densities. Later on, \cite{Ji:2019} ran 3D simulations of TRMLs to compare with previous analytic theories, finding surprisingly low inflow and turbulent velocities, as well as different scalings for the mixing layer width and surface brightness with cooling time. Recent TRML models based on high-resolution, three-dimensional simulations \citep{Fielding:2020, Tan:2021} have further elucidated the physical processes responsible for regulating the rate at which mass is exchanged through the fractal surface separating the hot and cold phases.

Besides three-dimensional simulations on a variety of scales that have yielded essential insights into the physics of TRMLs, another valuable approach to study TRMLs is through analytic, one-dimensional models. \cite{McKee:1977} were among the first to develop analytic solutions for the gas transfer between a spherical cloud and the surrounding hot gas. They modeled thermal conductivity as a power-law of temperature (taken from \cite{Massey:1975} and \cite{Spitzer:1962}) and identified a critical radius that determines whether the cold cloud evaporates or condenses. \cite{JGKim:2013} and \cite{Inoue:2006} both analyzed the structure and stability of TRMLs (which they called the transition layer that connects the warm and cold neutral medium) by finding one-dimensional, steady-state solutions to the fluid equations in the reference frame where the transition layer is stationary. \cite{Binette:2009} also developed a one-dimensional analytic model for TRML to understand its impact on observational signatures like line profile shapes. Notably, \cite{Binette:2009} adopted a mixing length approach similar to what we will discuss in \autoref{sec: cond and turb} and uses a constant turbulent viscosity. More recently, \cite{Tan:2021b} and \cite{Tan:2021} argued that understanding the competition between thermal conduction and cooling is a crucial step towards capturing the detailed phase structure of TRMLs, including the temperature distribution and column density. They demonstrated that despite its complex fractal structure, TRMLs can actually be described by a simple 1D conductive-cooling front model that quantitatively reproduces 3D simulation results including column densities and line ratios. These works suggest that studying TRMLs through 1D models is a promising approach that has the potential of reducing computational cost and helping us develop physical intuition. 

In this work, we develop an analytic, 1.5-dimensional\footnote{By 1.5-dimensional we mean that we include components of the velocity along and perpendicular to the spatial dimension.} model for Turbulent Radiative Mixing Layers (TRMLs). The crucial difference between our model and the models mentioned above is that we now explicitly include the impact of turbulence driven by the shear flow (which is likely to dominate over the physical conduction and viscosity in most regimes). We do this by introducing a simple parameterization of the effective turbulent conductivity and viscosity that is proportional to the shear velocity gradient. We also adopt a piece-wise power-law cooling curve that takes into account of both collisional ionization and photo-ionization equilibrium. Our model allows us to better quantify the role viscosity plays in TRMLs. It also reproduces the mass, energy, momentum transfer and shape of the phase distributions from 3D simulations \citep{Fielding:2020}. In \autoref{sec:methods}, we recast the steady state fluid equations into our desired form and introduce the cooling curve and effective turbulent conductivity and viscosity in our model, as well as the simulations we compare against. In \autoref{sec:results}, we present results of our model, including: the relationship between the mass flux $\dot m$ and total cooling $\Qcool$ derived from energy conservation, the relative importance of energy sources and sinks in TRMLs as a function of parameters of our model, a demonstration that we match key features of our analytic model with 3D simulations, and a determination of the turbulent Prandtl number $\PR$. In \autoref{sec:discussion}, we compare with previous works and discuss observational implications.

\section{Methods} \label{sec:methods}

\begin{figure*}
\centering
\includegraphics[width=\textwidth]{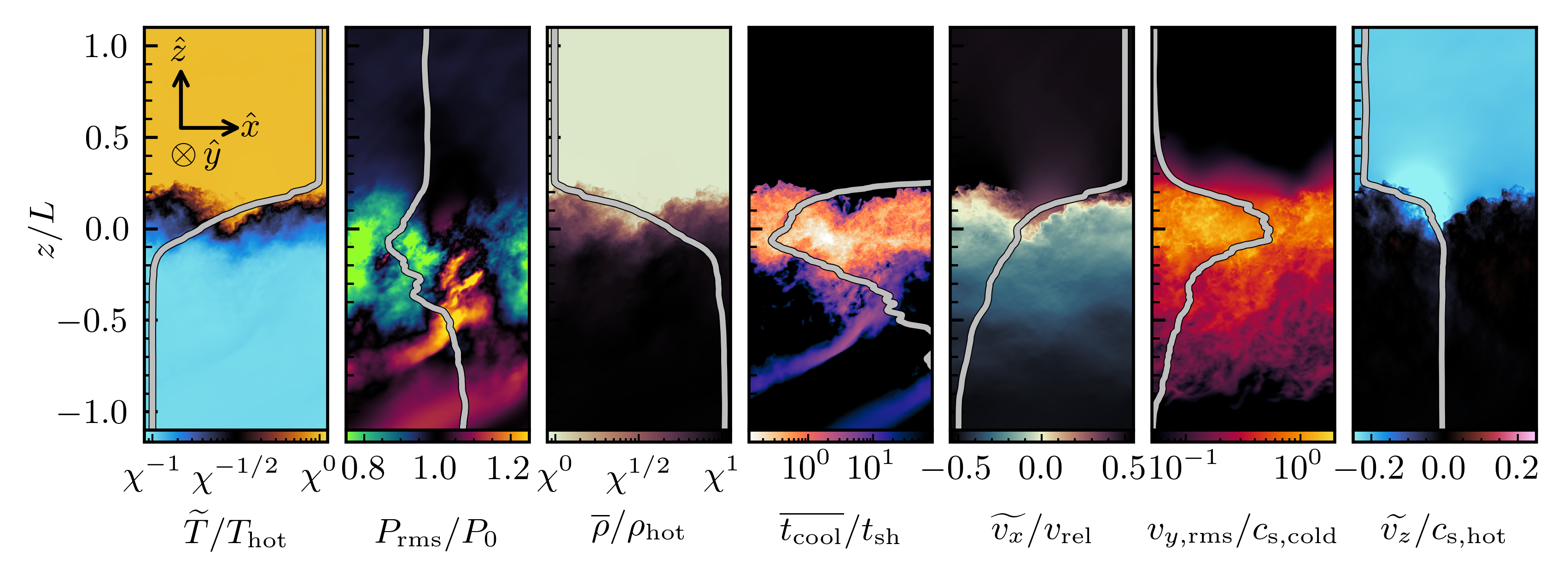} 
\caption{Example of the structure of a turbulent radiative mixing layer (TRML) three dimensional simulation similar to what was published by \cite{Fielding:2020} but using the cooling curve we have adopted here. Going from left to right the panels show the mass weighted average temperature $\widetilde T$, the root mean square pressure $P_{\rm rms}$, the volume weighted density $\overline{\rho}$, the average cooling time $\overline{t_{\rm cool}}$ (defined to be the ratio of integral of the thermal energy divided by the integral of the radiative cooling along the $\hat{y}$ direction), the mass weighted average $\hat{x}$-velocity $\widetilde{v_x}$, the root mean square $\hat{y}$-velocity $v_{y, {\rm rms}}$, and the mass weighted average $\hat{z}$-velocity $\widetilde{v_z}$. Overplotted in grey is the horizontally averaged profile of each of these quantities. This demonstrates the general configuration in which the hot phase is flowing in toward the cold phase and undergoes a rapid transition from hot to cold, which is where the vast majority of the cooling occurs. This is also the locus of the turbulence, which is traced out by $v_{y, {\rm rms}}$. The shear velocity $\widetilde{v_x}$ has a thicker transition region. }
\label{fig:schematic}
\end{figure*}

We consider a 1.5 dimensional model for a turbulent radiative mixing layer. The coordinate system we will adopt has the mass flux $\dot m$ between the phases in the $\hat{z}$ direction and the shear motion between the phases in the $\hat{x}$ direction. We will adopt a convenient notation in which the pressure $P = \rho T$, where we have absorbed the constant factor $\left. k_B \right/ (\mu m_p)$ into our definition of $T$ (with this convention $T$ has units of velocity squared, and can be thought of as the isothermal sound speed squared). Since this is a 1.5 dimensional model the derivatives perpendicular to $\hat{z}$ are all zero.

\begin{figure}
\centering
\includegraphics[width=\columnwidth]{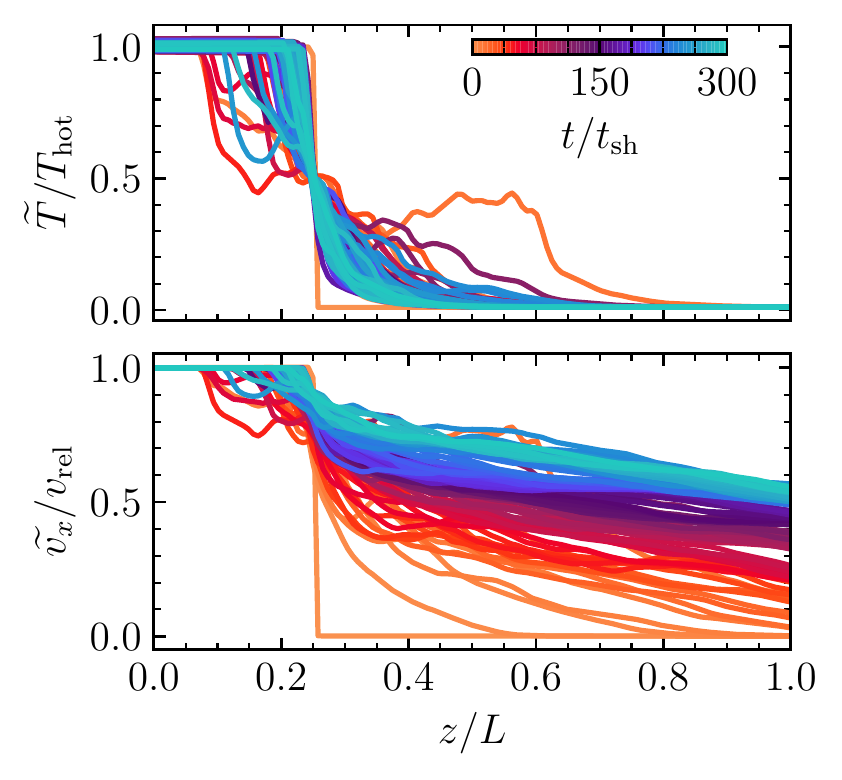} 
\caption{Time evolution of the horizontally averaged profiles of temperature (top) and shear velocity (bottom) weighted by mass in the same 3D turbulent radiative mixing layer (TRML) shown in \autoref{fig:schematic}. At each time the profiles are aligned to the $z$ location where $T = T_{\rm hot}/2$, which for this simulation corresponds to boosting into a reference frame shifted by a small $z$ velocity equal to $-\vrel/200$. While the shear velocity profiles continues to broaden in time the temperature profile quickly reaches an apparent equilibrium configuration and maintains it throughout the long duration of this simulation.}
\label{fig:aligned simulation profiles}
\end{figure}

The equations of mass, momentum, and energy conservation are
\begin{align}
    &\frac{\partial \rho}{\partial t} + \del \cdot \lr{ \rho \vv} = 0 \\
    &\frac{\partial \left(\rho \vv \right)}{\partial t} + \del \cdot \lr{ \rho \vv \vv + P} = \del \cdot \Tv
\end{align}
\vspace{-15pt}
\begin{align}
    \frac{\partial E}{\partial t} &+ \del \cdot \lr{ \vv (E+P)}\nonumber\\ 
    &= - \del \cdot \Qc + \del \cdot\lr{\Tv \cdot \vv} - \Edotcool ,
\end{align}
where the total internal energy is given by $E~=~1/2\rho v^2 + P/(\gamma-1)$, $\Tv~=~\mu\lr{\del\vv{+}\del\vv^\intercal{-}{2}/{3} \mathbb{I} \del\cdot\vv} $ is the viscous stress tensor, $\Qc = -\kappa \del T$ is the conductive heat flux, and $\Edotcool$ is the cooling rate. In reality, the conductivity $\kappa$ and viscosity $\mu$ of a fluid are mediated by the discrete collisions of the particles that comprise the fluid. In practice, in addition to the physical $\kappa$ and $\mu$, turbulence can effectively act as an additional form of ``eddy'' conductivity and viscosity, as has long been studied \citep[e.g.][]{Frisch:1995}. In this paper, we assume that the physical $\kappa$ and $\mu$ values are negligible and that the properties of TRMLs are dominated by the turbulent $\kappa$ and $\mu$. We will discuss how we model $\mu$ and $\kappa$ in more detail in \autoref{sec: cond and turb}.

Throughout this paper, we use 3D simulations of TRMLs as a point of comparison to our analytic model. The 3D simulations are described in detail in \autoref{sec:simulation description}. For now, we utilize two pieces of information from the 3D simulations to facilitate our derivation of the fluid equations. First, we show an annotated 2D projection from the 3D simulations in \autoref{fig:schematic} that clearly illustrates the basic geometry of the system we are dealing with. Second, we make use of a fundamental property of rapidly cooling TRMLs that has been uncovered in 3D simulations. Namely, over time the horizontally averaged shear velocity and momentum profiles continually expand due to the effective turbulent viscosity. The horizontally averaged temperature and inflow velocity profiles, however, do not. Instead these profiles quickly reach an apparent equilibrium configuration and maintain it indefinitely. An example of this behavior is shown in \autoref{fig:aligned simulation profiles} for the same 3D simulation shown in \autoref{fig:schematic}. Physically, this can be understood by comparing the timescales over which these properties change. The shear profile spreads out on a timescale proportional to the shear time $t_{\rm sh}~=~L/\vrel$, where $L$ is the characteristic lengthscale in the direction of the shear flow. The temperature and inflow velocity, on the other hand, evolve on a timescale proportional to the cooling time $\tcool$. Therefore, in the rapid cooling limit in which $\tcool \ll \tsh$, the temperature and inflow velocity profiles evolve much more rapidly than the shear velocity and, thus, can, to good approximation, be considered to have ample time to reach a quasi-steady state equilibrium configuration. Note that in the slowly cooling limit ($\tcool \gtrsim \tsh$), which we do not consider here, this assumption breaks down and all of the quantities will evolve on similar timescales.

In our model we make use of this quasi-steady state nature of all horizontally averaged quantities except the shear velocity to explicitly assume that the time derivative of $\rho$, $v_z$, and $T$ is zero, and only the time derivative of $v_x$ is non-zero.  Using this we can now expand the 4 equations that fully describe our system.
\begin{align}
    &\frac{\partial }{\partial z}{}\lr{\rho \vz} = 0 
\end{align}
\begin{align}
    \dt{}\lr{\rho \vx} + \frac{\partial }{\partial z}{}\lr{\rho \vx \vz} &= \frac{\partial }{\partial z}{}\lr{\mu \frac{\partial v_x}{\partial z}} \nonumber\\
    &= \mu \frac{\partial^2 v_x }{\partial z^2} + \frac{\partial \mu}{\partial z} \frac{\partial v_x}{\partial z}
\end{align}
\vspace{-15pt}
\begin{align}
    \frac{\partial }{\partial z}{}\lr{\rho \vz^2 + P} = \frac{\partial }{\partial z}{}\lr{\frac{4}{3} \mu \frac{\partial v_z}{\partial z}}
\end{align}
\begin{align}
    \dt{}\lr{\frac{1}{2} \rho \vx^2} + &\frac{\partial }{\partial z}{}\biggl(\rho \vz \lr{\frac{1}{2} \lr{\vx^2 + \vz^2} + \frac{\gamma}{\gamma-1} T} \nonumber\\
    &- \kappa \frac{\partial T}{\partial z} - \lr{\mu \lr {\vx \frac{\partial v_x}{\partial z} + \frac{4}{3} \vz \frac{\partial v_z}{\partial z}}} \biggr) \nonumber\\
    &= -\Edotcool
\end{align}

Which then gives us
\begin{align}
    &\dot m \equiv \rho \vz = \text{constant}  \\
    &\dot m \lr{ \frac{1}{\vz}\frac{\partial v_x}{\partial t} + \frac{\partial v_x}{\partial z}} = \frac{\partial }{\partial z}{} \lr{\mu \frac{\partial v_x}{\partial z}} \label{eq:partial derivatives 1}\\
    &\dot m \frac{\partial }{\partial z}{}\lr{ \vz + \frac{T}{\vz}} = \frac{\partial }{\partial z}{}\lr{\frac{4}{3} \mu \frac{\partial v_z}{\partial z}} \label{eq:partial derivatives 2}\\
    &\dot m \vx \frac{1}{\vz}\frac{\partial v_x}{\partial t} + \frac{\partial }{\partial z}{}\biggl(\dot m \lr{\frac{1}{2} \lr{\vx^2 + \vz^2} + \frac{\gamma}{\gamma-1} T} \nonumber \\
    &- \kappa \frac{\partial T}{\partial z} - \lr{\mu \lr {\vx \frac{\partial v_x}{\partial z} + \frac{4}{3} \vz \frac{\partial v_z}{\partial z}}} \biggr) = -\Edotcool \label{eq:partial derivatives 3},
\end{align}
where $\kappa$ is the conductivity, $\mu$ is the viscosity given by $\mu = \PR \kappa$, and $\Edotcool$ is the radiative cooling function.

At this point we are at a crossroads for how to solve the above system of equations. In principle, one could solve the system numerically as it is, which involves dealing with both spatial and temporal derivatives. Instead, we rely on the fact that all the time derivatives are applied on $v_x$, which is slowly varying. We exploit this feature to solve for a snapshot of the system in time by prescribing a profile for $v_x$ by hand. The details of how we do that will be described in \autoref{sec:cosine vx}. For now, we rearrange the above system to solve for expressions for the second derivatives of $\vz$ and $T$. The general form for these expressions for some arbitrary choice of $\kappa$ and a constant $\PR$ is
\begin{align}
    \ddvz &= \frac{3}{4} \frac{\dot m}{\mu} \lr{\frac{1}{\vz} \dT {+} \dvz \lr{1 {-} \frac{T}{\vz^2}}} {-} \dvz \frac{1}{\mu} \dz{\mu} \label{eq:z equation}
\end{align}
\vspace{-15pt}
\begin{align}
    \ddT &= \frac{\Edotcool}{\kappa} {+} \frac{\dot m \vz}{\kappa} \lr{\frac{T}{\vz^2} \dvz {+} \frac{1}{\gamma{-}1} \frac{1}{\vz} \dT} {-} \dT \frac{1}{\kappa} \dz{\kappa} \nonumber\\
    &\qquad \quad{-} \PR \lr{\lr{\dvx}^2 {+} \frac{4}{3} \lr{\dvz}^2}. \label{eq:T equation}
\end{align}
Since with our assumptions $T$ and $\vz$ are now only functions of $z$, we change all partial derivatives to full derivatives to emphasize that we are solving for a time-independent slice of the solution. Note that for the $\left. d v_x \right/ dz$ term that appears in \autoref{eq:T equation}, we plug in our prescribed $v_x$ profile, which is informed by results from 3D simulations. This precludes the need of simultaneously solving a third $\vx$ equation that involves time derivatives.

With the system of \autoref{eq:z equation} and \autoref{eq:T equation}, we can easily plug in different choices for $\kappa$ and $\PR$, and then solve the resulting equations numerically.

It is worth clarifying that the time-independence of \autoref{eq:z equation}, \autoref{eq:T equation}, and our prescribed $v_x$ profile does not mean that the evolution of TRMLs is time-independent. Solving \autoref{eq:z equation} and \autoref{eq:T equation} only gives us a ``snapshot'' of the TRML in time. To solve for the time evolution, one would need to find $\left. \partial v_x \right/ \partial t$ using
\begin{align}
    \frac{\partial v_x}{\partial t}  = \frac{\mu v_z}{\dot m} \frac{\partial^2 v_x }{\partial z^2} - \frac{\mu v_z}{\dot m} \frac{\partial v_x}{\partial z} \lr{ \frac{\dot m}{\mu} - \frac{1}{\mu} \frac{\partial \mu}{\partial z}} \label{eq:vx dot equation}
\end{align}
which comes from solving for the second derivative of $\vx$ in \autoref{eq:partial derivatives 1}, \autoref{eq:partial derivatives 2}, and \autoref{eq:partial derivatives 3}. Using \autoref{eq:vx dot equation} and an appropriate time step, one can time-evolve the original prescribed $v_x$ profile and plug the new $v_x$ profile into the system of \autoref{eq:z equation} and \autoref{eq:T equation} to solve for a new snapshot of the TRML at a later time. This process is akin to the approach that has been adopted for many other systems with multi-scale processes, such as stellar structure \citep[e.g.,][]{1990sse..book.....K}.

Turning our attention back to the system we are solving (\autoref{eq:z equation} and \autoref{eq:T equation}), it is instructive to rearrange \autoref{eq:T equation} into a more intuitive form:
\begin{align}
    &\overbrace{\color{black} \frac{d}{dz} \biggl(\frac{\gamma}{\gamma-1} \dot m T\biggr)}^{\color{enth_color} \rm advected \, enthalpy} - \overbrace{\color{black}\dz{}\biggl(\kappa \dT\biggr)}^{\color{cond_color} \rm conduction} = - \overbrace{\color{black}\Edotcool}^{\color{cool_color} \rm cooling} \nonumber\\
    &\qquad + \begingroup \color{xvis_color} \underbrace{\color{black}\mu \lr{\dvx}^2}_{\color{xvis_color} x-{\rm viscous\, heat}}\endgroup + \begingroup \color{zvis_color}\underbrace{ \color{black} \frac{4}{3} \mu \lr{\dvz}^2}_{\color{zvis_color} z-{\rm viscous\, heat}} \endgroup + \begingroup \color{work_color}\underbrace{\color{black}\vz \dz{P}}_{\color{work_color} \rm work} \endgroup , \label{eq: before integration}
\end{align}
where we have labeled each term by its physical meaning using the same colors that will be used in subsequent figures. The work term comes from the following relation 
\begin{align}
    \vz \dz{P} = \dot m \dz{T} - \dot m \frac{T}{vz} \dz{\vz} .
\end{align}

If we integrate this from hot to cold then we get
\begin{align} \label{eqn:sources and sink}
    \dot m &\lr{\frac{\gamma}{\gamma-1} \Delta T} = -\Qcool + \Hvisc + \mathcal{W} \\
    &\text{where  }  \Qcool = \int \Edotcool dz \label{eq:Qcool}\\
    &\Hvisc = \int \mu \lr { \lr{\dvx}^2 + \frac{4}{3} \lr{\dvz}^2} dz \\
    &\mathcal{W} = \int \vz \dz{P} dz .
\end{align}
Recall that we have absorbed the constants into our definition of $T$, so $\gamma \Delta T = c_{\rm s, hot}^2 - c_{\rm s, cold}^2 = \Delta c_{\rm s}^2$.

It is useful to think of \autoref{eqn:sources and sink} as a statement of energy sources and sinks. It tells us that the advected enthalpy from the hot phase (LHS of \autoref{eqn:sources and sink}) receives three contributions (RHS of \autoref{eqn:sources and sink}): radiated away by cooling, gained through viscous dissipation of the relative kinetic energy between the phases, or work done by the pressure gradient. 

\subsection{Conventions and Unit Conversion}
\label{sec:unit conversion}

In the remainder of this paper, many physical quantities will be expressed as dimensionless ratios. This keeps our approach as general as possible, but it can be useful to provide physical conversions to guide intuition and comparison to observations. In this section, we explicitly state the conventions we are using and discuss how to convert between code units and physical units.

\subsubsection{Dimensionless Numbers}
The structure of a TRML is strongly affected by three dimensionless numbers that characterize the physics of the system, which are
\label{sec:dimensionless numbers}
\begin{align}
    &\chi = \frac{T_{\rm hot}}{T_{\rm cold}}\\
    &\mathcal{M}_{\rm rel} = \frac{v_{\rm rel}}{c_{\rm s,hot}} = v_{\rm rel} \left( \frac{\gamma P_{\rm hot}}{\rho_{\rm hot}} \right)^{-\frac{1}{2}}\\
    &\tau = \frac{t_{\rm cool,min}}{t_0} = \frac{t_{\rm cool,min}}{\left. L_0 \right/ v_0} = \frac{t_{\rm cool,min}}{L_0} \left(\frac{k_B T_{\rm hot}}{\mu m_p}\right)^{\frac{1}{2}} ,
\end{align}
where $v_{\rm rel}$ is the relative velocity between the hot and cold phases, $c_{\rm s,hot}$ is the hot phase sound speed, $\gamma$ is the adiabatic index that we take to be 5/3 throughout, $t_{\rm cool,min}$ is the minimum cooling time that occurs at intermediate temperatures, $L_0$, $t_0$, and $v_0$ are the characteristic length, time, and velocity of the system, $\mu$ is the mean molecular weight and is equal to 0.62 in an ionized medium with solar metallicity, and $m_p$ is the mass of a proton. 

We set these characteristic scales by setting (i) the hot phase pressure $P_{\rm hot}$, (ii) the hot phase temperature $T_{\rm hot}$, and (iii) the minimum cooling time $t_{\rm cool, min}$. With these the velocity scale is $v_0 = \sqrt{\left. P_{\rm hot} \right/ \rho_{\rm hot}} = \sqrt{\left. k_B T_{\rm hot} \right/ \mu m_p}$, which differs from $c_{\rm s,hot}$ by $\sqrt{\gamma}$.

As we will see later in \autoref{sec:cooling curve}, the dimensionless number $\tau$ serves as a normalization to the cooling rate and the cooling time. In 3D simulations, this normalization is set relative to the shear time $t_{\rm sh} = \left. L \right/ v_{\rm rel}$, where $L$ is the shear layer length, which is well-defined in 3D TRML simulations. Thus by setting a physical scale for $L$, and the ratio $\tcool/t_{\rm sh}$, the physical cooling rate is determined. However, here in our 1.5D model we do not have a transverse lengthscale, which precludes us from implementing the same setup. We, therefore, take the opposite approach and set the physical timescale which in turn sets the physical lengthscale. We set the physical timescale using the following relation: 
\begin{align}
     t_{\rm cool,min} &= t_0 \tau \nonumber\\
     &= 0.131  \, {\rm Myr} \,  \left(\frac{\left. P \right/ k_B}{10^3 {\rm K cm^{-3}}}\right)^{-1}.
     \label{eq:t_cool_min and tau}
\end{align}
The numerical value in \autoref{eq:t_cool_min and tau} comes from using an approximate fit to the cooling rates from \cite{Wiersma:2009} assuming solar metallicity. In reality there should be a metallicity dependence on the minimum cooling time, however, because hydrogen is the primary coolant at the temperatures where the cooling time is minimized ($\sim 2\times10^4$ K ) the metallicity dependence is very weak, so we leave it out for sake of simplicity. 

Thus, at a fixed pressure, adjusting $\tau$ implies adjusting the characteristic time $t_0$, which is related to the characteristic length of the system by
\begin{align}
     L_0 &= v_0 t_0 = \left(\frac{k_B T_{\rm hot}}{\mu m_p}\right)^{\frac{1}{2}} t_0 \nonumber\\
     &= 378.4  \,{\rm pc} \, \left(\frac{\tau}{10^{-1.5}}\right)^{-1} \left(\frac{\left. P \right/ k_B}{10^3 {\rm K cm^{-3}}}\right)^{-1} \nonumber \\ 
     & \qquad  \qquad \qquad  \qquad  \left(\frac{T_{\rm hot}}{10^6 {\rm K}}\right)^{\frac{1}{2}},
     \label{eq: lengthscale}
\end{align}
where $v_0$ is the characteristic velocity, and $t_0$ is the characteristic timescale that comes from \autoref{eq:t_cool_min and tau}.

In other words, by adjusting $\tau$ we are effectively adjusting the lengthscale of the system. This characteristic length $L_0$ allows us to convert other lengthscales in the problem into physical units. For example, the thickness of the mixing layer, which we will later denote as $h$, is given by $h \times L_0$ in physical units.

Whenever possible, we also express physical quantities in a dimensionless form. For example, we speak in terms of the dimensionless $\left. T \right/ T_{\rm hot}$ instead of $T$. We can trivially make the conversion to dimension-full quantities by multiplying through by the hot phase temperature $T_{\rm hot}$, which can be chosen freely based on the physical scenario at hand. The exact same logic holds for pressure (which we normalize by the hot phase pressure $P_{\rm hot}$) and density (which can be calculated from temperature and pressure using the ideal gas law).

\subsubsection{The Phase Distributions}
\label{sec:phase distributions}
To understand the detailed phase structure of TRMLs, it is instructive to examine the distribution of mass and the amount of cooling as a function of temperature. We call these distributions the phase distributions. Here we discuss the definition and calculation of the cooling flux and column density distribution of TRMLs.

The cooling flux distribution, or $\left. d F_{\rm cool} \right/ d {\rm logT}$, tells us how the cooling flux is distributed in log temperature space. The cooling flux distribution can be calculated by
\begin{align}
    \frac{d F_{\rm cool}}{d {\rm logT}} = \frac{T}{\frac{dT}{dz}} F_{\rm cool} = \frac{T}{\frac{dT}{dz}} \frac{\Edotcool(z)}{L_0^2},
\end{align}
where $\Edotcool(z)$ is rate of energy loss from radiative cooling.

Similarly, the temperature distribution of mass per unit area, or more conveniently column density $N$, is given by
\begin{align}
    \frac{d N}{d {\rm logT}} = \frac{d}{d {\rm logT}} \left(\frac{M}{L_0^2 m_H}\right) = \frac{T}{\frac{dT}{dz}} \frac{\rho}{L_0^2 m_H} , \label{eq:dN_dlogT}
\end{align}
where $m_H$ is the mass of a hydrogen atom.

The cooling flux and column density distributions are also normalized to be dimensionless as follows:
\begin{align}
    \frac{1}{F_0} \frac{d F_{\rm cool}}{d {\rm logT}} &\text{    where    } F_0 = \frac{\rho_{\rm hot} c_{\rm s,hot}^3}{L_0^2} ; \\
    \frac{1}{N_0} \frac{d N}{d {\rm logT}} &\text{    where    } N_0 = \frac{\rho_{\rm hot} L_0}{m_H} .
\end{align}

\subsection{Cooling Curve}
\label{sec:cooling curve}
Since we are interested in general properties and behavior of TRML, we adopt a simplified cooling rate that depends only on temperature; the overall shape is intended to be similar to more realistic curves but is both simpler and customizable. In particular, we use a cooling curve that is a piece-wise power-law. The power-law slopes are $\betalo = -2$ if $ T \leq \Tpeak$ and $\betahi = 1$ if $T > \Tpeak$. Additionally, we include a heating term such that the cooling curve has two equilibria at both the cold and the hot phase temperatures. In realistic systems the hot phase does not have a stable equilibrium and is instead very slowly cooling. We adopt this artificial hot phase equilibrium to improve numerical performance but in practice it has no appreciable impact on our results. We discuss the cooling curve we use in more quantitative details in appendix.

In \autoref{fig:cooling time comparison}, we compare the cooling time profile obtained from our piece-wise power-law cooling curve with the realistic cooling time profile. For most of the remainder of this paper, we are going to use our piece-wise power-law cooling curve as a proxy to the realistic cooling curve. However, in \autoref{sec:links to observations, column density} and \autoref{sec:links to observations, surface brightness} where we calculate the ion column density and emission line surface brightness of TRMLs from results of our analytic solutions, we use the realistic cooling curve to obtain more accurate results.

\begin{figure}
\centering
\includegraphics[width=\columnwidth]{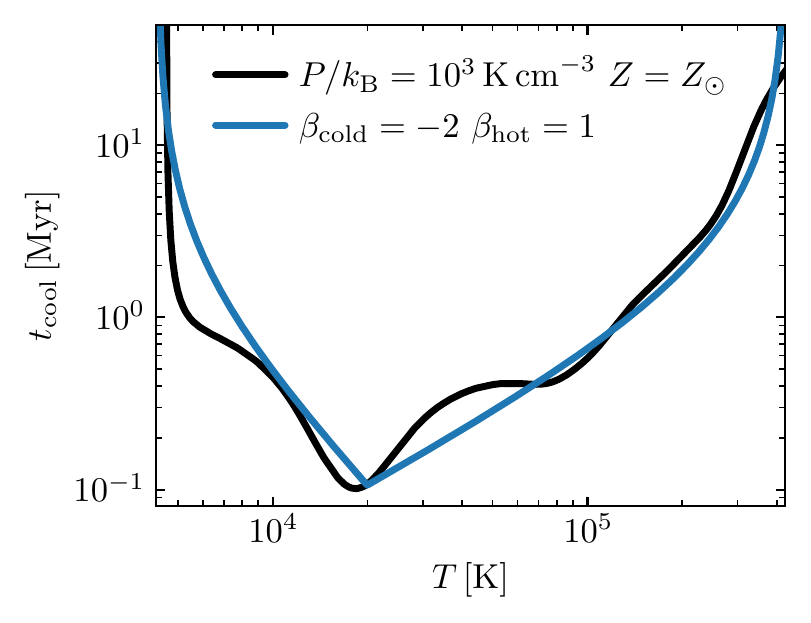} 
\caption{Comparison of the realistic cooling time profile with our piece-wise power-law cooling time prescription. The realistic cooling time profile (in black) assumes $\left. P \right/ k_b = 10^3 {\rm K} {\rm cm}^{-3}$ and solar metallicity. The piece-wise power-law profile (in blue) has $\betalo=-2$ and $\betahi=1$, where $\betalo$ and $\betahi$ are the power-law slopes at $T \leq \Tpeak$ and $T > \Tpeak$, respectively. We discuss our piece-wise power-law prescription in more details in \autoref{sec:cooling curve}. As shown in the figure, this prescription closely resembles the realistic profile and has the advantage of being simpler and easily customizable.}
\label{fig:cooling time comparison}
\end{figure}

\subsection{Conductivity \& Viscosity}
\label{sec: cond and turb}

\cite{Smagorinsky:1963} is one of the first to develop a framework for modeling viscosity from turbulence in eddies. In his pioneering subgrid-scale model for large eddy simulations, \cite{Smagorinsky:1963} defined what he called the eddy viscosity as

\begin{align}
    \mu_{\rm t} = \ell^2 \sqrt{ \bar{S_{\rm ij}} \bar{S_{\rm ij}}}
    \label{eq:Smagorinsky viscosity}
\end{align}
where $\ell$ is the lengthscale of the large eddies, $\bar{S_{\rm ij}}$ is the strain rate tensor, and $\sqrt{ \bar{S_{\rm ij}} \bar{S_{\rm ij}}}$ gives the velocity scale corresponding to the lengthscale $\ell$.

\cite{Prandtl:1925} also took the large-scale eddy viscosity (comparable to what we will later refer to as the effective turbulent viscosity in our analytic model) into account in his mixing length theory. He recognized this eddy viscosity as characterizing the transport and dissipation of momentum and energy due to turbulence on small scales. Ultimately, the physics of turbulence and viscosity determines the distance a gas cloud can travel with its original temperature and density before blending into the surrounding. \cite{Prandtl:1925} called this distance ``the mixing length". 

Mixing length theory \citep{Prandtl:1925} has been used in previous works on TRMLs. For instance, \cite{Tan:2021} developed a 1D mixing length model of TRMLs (with a constant or temperature-dependent conductivity and no viscosity) and computed the resulting mean density and temperature profiles that matches well with 3D simulation results. 

Motivated by \cite{Smagorinsky:1963} and \cite{Prandtl:1925}, we introduce a model for effective turbulent conductivity and viscosity. The conductivity is given by 
\begin{align}\label{eq:conductivity}
    \kappa = f_{\nu} h^2 \rho \left| \dvx \right| + \kappa_0 ,
\end{align}
where $\rho$ and $h$ are the density and thickness of the mixing layer, $v_x$ is the shear velocity, and $\kappa_0$ is a small additive constant used to prevent singularities when $\dvx \rightarrow 0$. The $\kappa_0$ term is somewhat analogous to the numerical dissipation in the 3D simulations and we find that as long as it is small compared to the first term our results are insensitive to the exact choice (we adopt $\kappa_0 = 10^{-6}$ throughout). We include the absolute value on $\left. dv_x \right/ dz$ because $\kappa$ must always be greater than 0. Note that in the definition of $\kappa$ we introduce the constant, dimensionless prefactor $f_{\nu}$, which can be used to adjust the magnitude of $\kappa$. Another useful way to understand the utility of $f_{\nu}$ is to examine the combination $f_{\nu} h^2$, which has dimensions of length squared. By defining $f_{\nu}$ we are effectively introducing a new lengthscale given by $\left( f_{\nu} h^2 \right)^{1/2}$ $ = f_{\nu}^{1/2} h$ that is equivalent to $\ell$ in the Smagorinsky/mixing length formalism. Since in our 1.5D model we do not have the freedom to set a transverse lengthscale as in the 3D simulation, this new lengthscale defined by $f_{\nu}$ is introduced to compensate for the lost of that degree of freedom. 

We assume that the effective turbulent viscosity has the same functional form as the conductivity, and is thus defined as
\begin{align}
    \mu = \PR \kappa = \PR f_{\nu} h^2 \rho \left| \dvx \right| + \PR \kappa_0,
\end{align}
where $\PR$ is the effective turbulent Prandtl number.

We note that in previous works on analytic, 1D models of TRMLs \citep[e.g.,][]{Tan:2021b}, the conductivity $\kappa$ is frequently modelled as either a constant or a function of temperature. This temperature dependence is either taken to follow the Spitzer conductivity scaling ($\kappa \propto T^{5/2}$) \citep{Spitzer:1962}, or is given an arbitrary power law scaling that is calibrated to match the results of 3D simulations. In these models the viscosity $\mu$ is not taken into account. In this way, our setup is fundamentally different from these previous works as here we compute the effective turbulent conductivity and viscosity based on the shear velocity profile (see Section~\ref{sec:cosine vx}), which enables us to connect more closely to the underlying physical processes.

\subsection{Description of Benchmark 3D Simulations}\label{sec:simulation description}
In order to better motivate the choices inherent in our model and to calibrate the parameters we have introduced, $f_{\nu}$ and $\PR$, we compare to a series of 3D simulations. These simulations are similar to what were presented in \cite{Fielding:2020} with the primary difference being that they use the cooling curve as we have adopted here.  A full description of these simulations as well as a detailed analysis will be presented in an upcoming work. We now briefly summarize these simulations and their analysis.

The simulations were performed with the \texttt{athena++} code framework \citep{athena++}, which solves the standard fluid equations (see equations 1-3). The simulations were performed with no explicit conduction or viscosity. As mentioned, the simulations include the exact same cooling and heating as given in \autoref{eq:edot_cool}.

We now briefly describe the simulation setup and the suite of parameter variations, a more comprehensive explanation will be included in a forthcoming companion paper. The simulations are initialized in pressure equilibrium between the hot and cold phase with a smooth temperature transition spread out over 4 cells. The hot and cold phases are initialized so that they are moving relative to each other in the $\hat{x}$ direction with a velocity given by $\vrel$. As with the temperature, the shear velocity transition is resolved by 4 cells. All three components of the velocity at the interface between the hot and cold phase is perturbed in the initial conditions. The root-mean-square of the velocity perturbation at the interface is set to $\vrel/20$ and falls off exponentially with distance from the interface with a scale length of 16 cells. The perturbations are generated in Fourier space and have equal power on wave numbers $1 \leq k L / 2\pi \leq 16$. 

Using this setup we ran a suite of simulations varying $\chi = 10^{1.75}$ to $10^{2.75}$, $\Mach = 0.29$ to $1.46$, and $\tau = 10^{-2}$ to $10^{-0.75}$. Our fiducial simulation, with $\chi = 100$, $\Mach = 0.58$, and $\tau = 10^{-1.5}$, is shown in \autoref{fig:schematic}.

As the simulation progresses in time the shear velocity/momentum transition layer spreads. The temperature, density, and inflow velocity respond quickly to the gradual spreading of the shear velocity and maintain nearly the same horizontally averaged profile. This is shown explicitly in \autoref{fig:aligned simulation profiles}. This motivates our assumption that these quantities can be taken to be in steady state while the shear velocity cannot. We must, therefore, assume some shear velocity profile in order to solve our 1.5D model. We discuss our approach to doing so in the next subsection. We need to first, however, decide \emph{when} to measure the simulation properties since they evolve with time. In order to standardize the measurements across the range of parameters all 3D simulation properties are measured when the thickness of the mixing layer is equal to the transverse width of the box, i.e. $h = L$. \footnote{We have verified that the exact $h$ value does not introduce any qualitative changes to our findings, as long as it is standardized across all simulations.} We measure the thickness $h$ by calculating the distance between the places where the $\hat{x}$ and $\hat{y}$ mass-weighted average shear velocity $\widetilde{v_x}$ is within 5 percent of the initial values (i.e. between 0.05 and 0.95 $\vrel$). We tested the sensitivity to the exact definition and found only minor quantitative differences. Because this definition is somewhat imprecise in practice we take the time average of when $h$ is within 5 percent of $L$. Again the exact choice has no substantive impact on our results, but helps avoid fluke temporal fluctuations. When comparing to simulation properties in later sections we use error bars to denote these temporal fluctuations about the mean.

\subsection{Taking a Prescribed Shear Velocity ($v_x$) Profile from the 3D Simulation}
\label{sec:cosine vx}
Instead of solving \autoref{eq:vx dot equation}, which contains a time derivative of $v_x$, together with \autoref{eq:z equation} and \autoref{eq:T equation}, we resort to the 3D simulations to obtain a prescribed $v_x$ profile and solve only for \autoref{eq:z equation} and \autoref{eq:T equation}.

\begin{figure}
\centering
\includegraphics[width=\columnwidth]{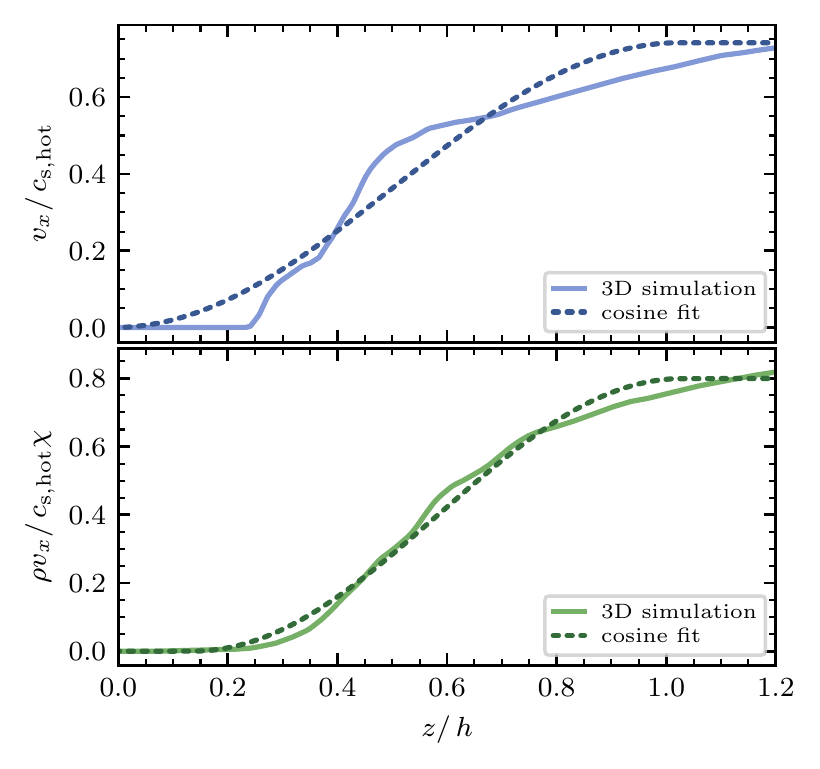} 
\caption{Comparison of the $v_x$ and $\rho v_x$ profiles from 3D simulation and the cosine fits that we use in our analytic model. The solid curve on the top panel is the mass weighted x-velocity from the 3D simulation, and the solid curve on the bottom panel is the volume weighted density times the mass weighted x-velocity from the 3D simulation, which gives us a $x$-momentum profile. The dotted curves on the top and bottom panels are the cosine fits to the $v_x$ and $\rho v_x$ profiles, respectively. We discussed these cosine fits in more details in \autoref{sec:vx profile setup} and \autoref{sec:momentum profile setup}.}
\label{fig:vx and momentum cosine fits}
\end{figure}

We explore two different ways of extracting the $v_x$ profile from 3D simulations. First, the most straightforward way is to directly take the mass weighted horizontally averaged $v_x$ profile from the 3D simulation. The physics of the system at hand dictates that the $v_x$ profile we use in our analytic model has to be continuous and perfectly flat at both the hot and cold phases ($\left. d v_x \right/ dz = 0$ at both ends of the solution.) Additionally, it is important to remember that the ultimate purpose of extracting the $v_x$ profile is to feed values of $\left. d v_x \right/ dz$ into our definition of the effective turbulent conductivity and viscosity in \autoref{sec: cond and turb}, which means whatever $v_x$ profile we come up with must also be differentiable. Due to these considerations, we choose to fit the 3D simulation $v_x$ profile with a cosine \footnote{All we are looking for in our $v_x$ profile is an approximate fit to the 3D simulation. Many other choices could have been used, but in experimenting, we found the cosine profile to be the simplest and produces useful results. Other possible choices like the logistic function or the hyperbolic tangent function produce qualitatively similar results, but we find that the non-zero gradient at the ends of these alternative choices affect the performance of our numerical integrator.}, as shown in the top panel of \autoref{fig:vx and momentum cosine fits}. We find the cosine fit convenient because it satisfies all the requirements mentioned above, has a simple functional form, and provides a reasonable fit all at the same time.

Alternatively, we also tried extracting the $x$-momentum profile $\rho v_x$ from 3D simulations and fit it with a cosine. Then, the $v_x$ profile we need can be obtained by dividing the $x$-momentum profile by $\rho$, or $\left. \dot m \right/ v_z$. Although this approach is more algebraically complex, one immediate benefit is that the cosine fit is much more suitable for the $x$-momentum profile, as shown in the bottom panel of \autoref{fig:vx and momentum cosine fits}. As we will see later, the $x$-momentum profile setup also exhibits some characteristics we see from the 3D simulations but not in the $v_x$ profile setup, which makes these two approaches complementary.

Next, we rigorously define the $v_x$ profile and $\rho v_x$ profile we are using and discuss the algebraic steps that need to be taken to simplify \autoref{eq:z equation} and \autoref{eq:T equation} in each of the two setups. In \autoref{sec:results}, we begin our investigation with the $v_x$ profile prescription and later compare results from the 3D simulations with results from both the $v_x$ and the $\rho v_x$ profiles.

\subsection{$v_x$ Profile as a Cosine}
\label{sec:vx profile setup}
Here we define $\vx(z)$ as
\begin{align}
    v_x(z) = 
\begin{cases}
    v_{\rm rel} & \text{if } z < 0 \\
    \frac{v_{\rm rel}}{2} \left(\cos\left(\frac{\pi z}{h}\right)+1\right) & \text{if } 0 < z < h \\
    0 & \text{if } z > h \\
\end{cases}
\end{align}
Consequently, if $0 < z < h$, we have
\begin{align}
    & \dvx = -\frac{v_{\rm rel}}{2} \frac{\pi}{h} \sin\left(\frac{\pi z}{h}\right) \label{eq:vx profile dvx}\\
    & \ddvx = -\frac{v_{\rm rel}}{2} \left(\frac{\pi}{h}\right)^2 \cos\left(\frac{\pi z}{h}\right) , \label{eq:vx profile d2vx}
\end{align}
where $h$ is the thickness of the mixing layer, and $v_{\rm rel} = |v_{x}(z=0) - v_{x}(z=h)|$. 

Using \autoref{eq:conductivity} and $\rho = \left. \dot{m} \right/ v_z$, we can express $\left. d \kappa \right/ dz$ as:
\begin{align}
    & \frac{d\kappa}{dz} = f_{\nu} h^2 \left(\frac{\dot m}{v_z} \frac{\dvx}{\left| \dvx \right|} \ddvx  - \frac{\dot m}{v_z^2} \dvz \left| \dvx \right|\right) . \label{eq:vx profile dkappa}
\end{align}
Plugging \autoref{eq:conductivity} and \autoref{eq:vx profile dkappa} into \autoref{eq:z equation} and \autoref{eq:T equation} gives us a system we can numerically integrate.

\subsection{$\rho v_x$ Profile as a Cosine}
\label{sec:momentum profile setup}

Next, let's look at prescribing the $x$-momentum $\rho v_x$ as a cosine profile. Before defining the profile, we first note a crucial constraint due to the expression of the shear velocity ($v_x$) gradient. The shear velocity gradient is given by
\begin{align}
    \dvx &= \frac{d}{dz} \left(\frac{p_x}{\rho}\right) = \frac{d}{dz} \left(\frac{p_x v_z}{\dot m}\right) \nonumber\\
    &= \frac{1}{\dot m} \left(p_x \dvz + v_z \frac{d p_x}{dz}\right) ,
\end{align}
where $p_x = \rho v_x$ is the $x$-momentum profile (which we will later define as a cosine). The constraint here is that the two terms in the expression for $\left. d v_x \right/ dz$ must have the same signs. Otherwise, $\left. d v_x \right/ dz$ would evaluate to 0 somewhere in the solution, causing a singularity and thus difficulty for the integration to proceed. Since we know for sure that $p_x>0$, $\left. d v_z \right/ dz <0$, and $v_z>0$ throughout the solution, this implies that $ \left. d p_x \right/ dz$ must be less than 0. With this in mind, we define $p_x(z)$ as
\begin{align}
    p_x \left(z\right) = 
\begin{cases}
    \rho_{\rm hot} v_{\rm rel} & \text{if } z < 0 \\
    \frac{\rho_{\rm hot} v_{\rm rel}}{2} \left(\cos\left(\frac{\pi z}{h}\right)+1\right) & \text{if } 0 < z < h \\
    0 & \text{if } z > h \\
\end{cases}
\end{align}
where $\rho_{\rm hot} = \left. P_{\rm hot} \right/ T_{\rm hot}$ is a known boundary condition at the hot end (see \autoref{sec:integration details} for more details). 
Under this setup, we have
\begin{align}
    & \kappa = f_{\nu} h^2 \frac{\dot m}{v_z} \left| \dvx \right| + \kappa_0 \nonumber\\
    &= \frac{f_{\nu} h^2}{v_z} \left|p_x \dvz + v_z \frac{d p_x}{dz}\right| + \kappa_0 \label{eq:momentum profile kappa}
\end{align} 
\begin{align}
    \frac{d\kappa}{dz} = &- \frac{f_{\nu} h^2 \dot m}{v_z^2} \dvz \left|\dvx\right| \nonumber\\
    &+ \frac{f_{\nu} h^2}{v_z} \frac{\dvx}{\left|\dvx\right|} \left(v_z \frac{d^2 p_x}{dz^2} + 2 \frac{d p_x}{dz} \dvz + p_x \ddvz\right) .
     \label{eq:momentum profile dkappa}
\end{align}
Note that the last term of \autoref{eq:momentum profile dkappa} has a second derivative of $v_z$ in it, which means after we plug \autoref{eq:momentum profile kappa} and \autoref{eq:momentum profile dkappa} into \autoref{eq:z equation} and \autoref{eq:T equation}, we still need to rearrange and solve for the second derivatives in order to obtain a system that is numerically integrable. After some algebra, we get
\begin{align}
    \ddvz =&  \frac{1}{{\mu+\frac{\PR f_{\nu} h^2}{v_z} \frac{\dvx}{\left|\dvx\right|} p_x \dvz}}\nonumber\\
    &\biggl(\frac{3}{4} \dot m \left(\frac{1}{v_z} \dT + \dvz\left(1-\frac{T}{v_z^2}\right)\right)\nonumber\\
    &+ \frac{\PR f_{\nu} h^2 \dot m}{v_z^2} \left(\dvz\right)^2 \left|\dvx\right| \nonumber\\
    &- \frac{\PR f_{\nu} h^2}{v_z} \dvz \frac{\dvx}{\left|\dvx\right|} \left(v_z \frac{d^2 p_x}{dz^2} + 2 \frac{d p_x}{dz} \dvz\right)\biggr)
\end{align}
\vspace{-15pt}
\begin{align}
    \ddT =& \frac{1}{\kappa} \biggl(\frac{\dot m}{\gamma-1}\dT + \frac{\dot m T}{v_z}\dvz + \Edotcool \nonumber \\
    &- \mu \lr{\lr{\dvx}^2 + \frac{4}{3} \lr{\dvz}^2} - \frac{d\kappa}{dz} \dT\biggr) .
\end{align}

\subsection{Carrying Out the Numerical Integration}
\label{sec:bisection on mdot}
After prescribing the $v_x$ or the $x$-momentum profile, we are left to numerically integrate two coupled differential equations to solve for the z-velocity ($v_z$) and temperature profiles. We set appropriate initial conditions and use scipy's solve-ivp integrator to perform the numerical integration.

\begin{figure}
\centering
\includegraphics[width=\columnwidth]{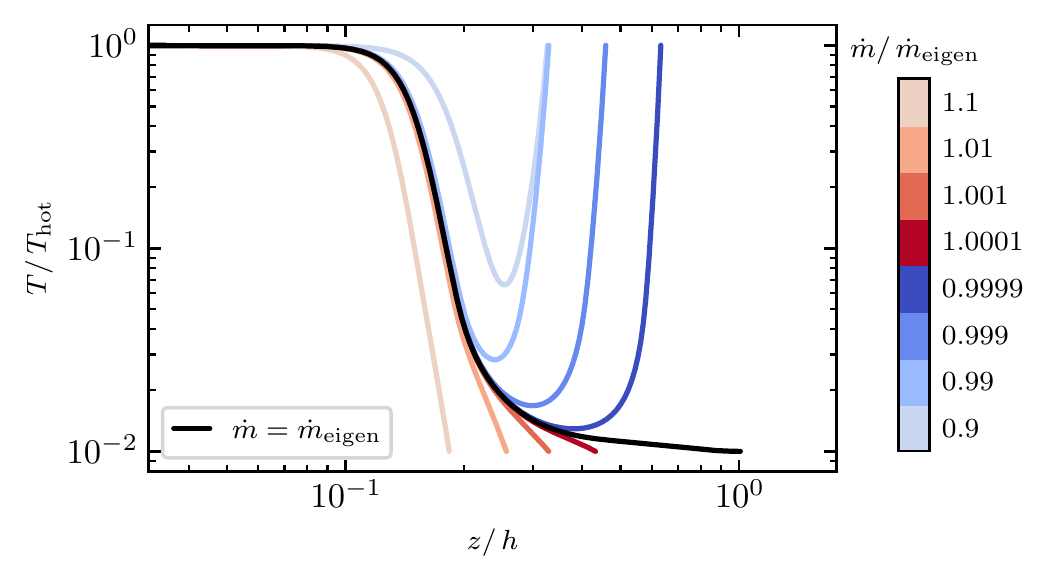}

\caption{There is only one value of the mass flux $\dot m$ that allows the temperature profile to reach the cold phase temperature smoothly ($\left. dT \right/ dz = 0$ at the cold phase). We call it the eigenvalue of $\dot m$ (${\dot m}_{\rm eigen}$). Here we plot 9 temperature profiles, varying the value of the mass flux $\dot m$ by $\pm10\%$ around ${\dot m}_{\rm eigen}$. The $T$ profile corresponding to ${\dot m}_{\rm eigen}$ is shown in black. When $\dot m > {\dot m}_{\rm eigen}$, the $T$ profile shoots down to the cold phase with a steep slope, with $\left. dT \right/ dz < 0$ at the end of the solution (the red group of solutions in the figure); when $\dot m < {\dot m}_{\rm eigen}$, the $T$ profile barely reaches the cold phase before bouncing back up, with $\left. dT \right/ dz > 0$ at the end of the solution (the blue group of solutions in the figure). The different signs of $\left. dT \right/ dz$ at the end of the solution in these two cases allow us to perform a bisection in $\dot m$ to pinpoint ${\dot m}_{\rm eigen}$. Our bisection method resolves ${\dot m}_{\rm eigen}$ to an accuracy of 1 part in $10^{15}$.}
\label{fig:T profiles at different mdot}
\end{figure}

An additional constraint of this problem is that the temperature profile needs to be flat ($\left. dT \right/ dz = 0$) at both the hot and cold phases by definition. For any given choice of parameters, there's only one value of the mass flux $\dot m$ that allows for that (as shown in \autoref{fig:T profiles at different mdot}). We call that the eigenvalue of $\dot m$. To find this eigenvalue, we bisect the parameter space of $\dot m$ and use scipy's root finder optimize.root-scalar to pinpoint the eigenvalue.

In \autoref{sec:integration details}, we provide more details on the numerical integration procedure.

\section{Results}\label{sec:results}

In this section, we present results from our analytic model of TRMLs. In \autoref{sec:vx profile individual sol}, we examine a fiducial solution of TRML produced by our analytic model. We discuss how changing the relative shear velocity changes the phase structure of TRMLs to help the readers build intuition on the physical processes at play. In \autoref{sec:Bernoulli flux} and \autoref{sec:energy sources and sinks vs.Pr and fnu}, we discuss the energy budgeting in TRMLs, identify the energy conservation criterion $\Hvisc = \left. \dot m v_{\rm rel}^2 \right/ 2$, and demonstrate how this criterion constrains the choices of model specific parameters $\PR$ and $f_{\nu}$. In \autoref{sec:vx profile, energy sources and sinks vs. vrel} and \autoref{sec:momentum profile, energy sources and sinks vs. vrel}, we examine how the relative importance of the energy sources and sinks in TRMLs change as a function of the relative shear velocity for the two variations of our analytic model (cosine $v_x$ profile and cosine $\rho v_x$ profile). Up until this point, all the physical insights are obtained purely through our analytic model. In \autoref{sec:matching mdot and Qcool}, \autoref{sec:vx, momentum, 3D, energy sources and sinks vs. vrel}, and \autoref{sec:matching phase distributions}, we discuss how to match the results from our analytic model with the 3D simulations, including matching $\dot m$, $\Qcool$, and the shapes of the phase distributions.

\subsection{Structure of TRMLs (with the Cosine $v_x$ Profile)}\label{sec:vx profile individual sol}

We first turn our attention to analyzing the detailed structure of solutions we obtained with the cosine $v_x$ profile prescription. We argue that the relative shear velocity ($\mathcal{M}_{\rm rel}$) between the hot and cold phase affects the solutions the most in our model. Varying the choice of $\mathcal{M}_{\rm rel}$ has two noteworthy effects. First, the $x$-component of viscous heating is proportional to $\left( \left. d v_x \right/ dz\right)^2$, which means larger $\mathcal{M}_{\rm rel}$ leads to more viscous heating. Second, as defined in \autoref{eq:conductivity}, the conductivity $\kappa$ is proportional to $\left. d v_x \right/ dz$, so larger $\mathcal{M}_{\rm rel}$ also implies higher conductivity. 

In what follows, we first analyze a solution with a fiducial choice of $\mathcal{M}_{\rm rel}$ to identify some key characteristics in the structure of the mixing layer. Next, we compare three solutions with low, medium and high values of $\mathcal{M}_{\rm rel}$ to understand how the previously identified characteristics change across different $\mathcal{M}_{\rm rel}$.

\subsubsection{A TRML with a Fiducial Choice of the Relative Shear Velocity $\mathcal{M}_{\rm rel}$}
\label{sec:fiducial vrel solution}

\begin{figure*}
\centering
\includegraphics[width=\textwidth]{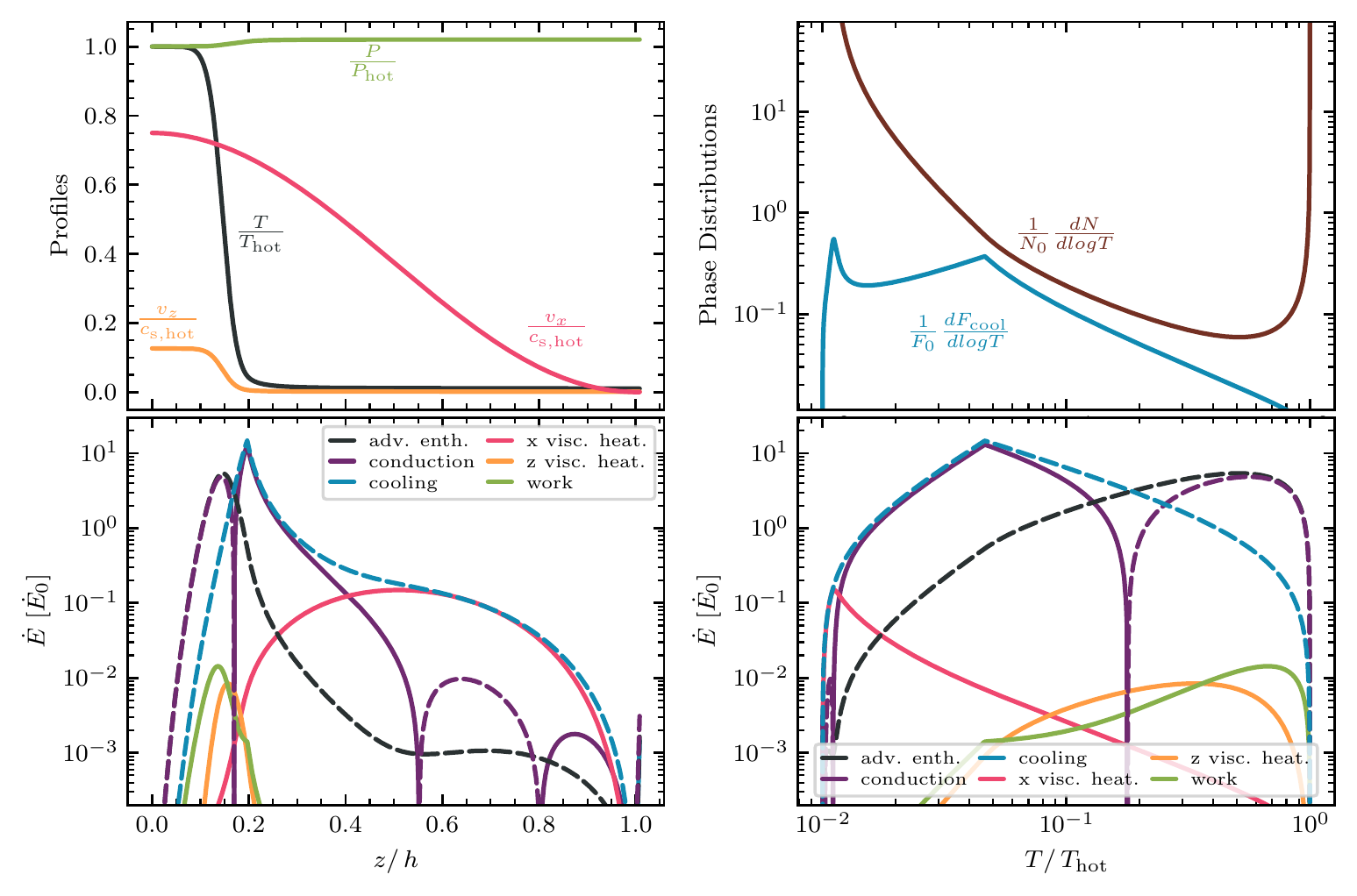}
\caption{A fiducial solution of our analytic model using the cosine $v_x$ profile. The parameter choices for this solution are: $\mathcal{M}_{\rm rel}=0.75$, $\chi=10^2$, $\tau=10^{-1}$, $f_{\nu}=10^{-2}$, $\PR=0.1$. The top left panel shows the temperature, pressure, and velocity profiles. The top right panel shows the cooling flux and column density distributions as a function of temperature, which are normalized by $F_0 = \rho_{\rm hot} c_{\rm s,hot}^3 / L_0^2$ and $N_0 = \left. \rho_{\rm hot} L_0 \right/ m_H$, respectively, where $\rho_{\rm hot}$ and $c_{\rm s,hot}$ are the hot phase density and sound speed, $L_0$ is the characteristic lengthscale of the mixing layer, and $m_H$ is the mass of a hydrogen atom. The bottom panels shows the various terms in the enegy flux (\autoref{eq: before integration}) versus position (left) and temperature (right), where solid lines indicate positive values and dotted lines indicate negative values. The black curves are related to temperature ($T$) and temperature gradients ($\left. dT \right/ dz$), including the temperature profile in the top left panel and the advected enthalpy in the bottom panels. The blue curves are related to cooling, including the cooling flux distribution in the top right panel and radiative cooling in the bottom panels. The green curves are related to pressure, including the pressure profile in the top left panel and the work done by the pressure gradient in the bottom panels. The pink and orange curves are related to the x- and z-velocity, respectively, including the $v_x$ and $v_z$ profiles in the top left panel and the x- and z-component of viscous heating in the bottom panels. Starting from the hot phase, this solution is initially dominated by conductive cooling. Then radiative cooling becomes important at intermediate temperatures to balance conductive heating. Finally, viscous heating takes over conductive heating to balance with radiative cooling near the cold phase. Notably, the region where viscous heating is important takes up a tiny fraction of the temperature space but a significant portion of position space.}
\label{fig:vx profile example sol}
\end{figure*}

In this section, we analyze the structure a mixing layer with $\mathcal{M}_{\rm rel}=0.75$.

The first step towards understanding the mixing layer is to look at the behavior of the key physical quantities that are involved. In our case, we are most interested in the temperature ($T$), velocity ($v_z$), shear velocity ($v_x$), and pressure ($P$) within the mixing layer. The top left panel of \autoref{fig:vx profile example sol} shows these 4 quantities plotted against the vertical distance into the mixing layer. To begin with, we note that the shear velocity ($v_x$) profile is prescribed to be a cosine function, as discussed in \autoref{sec:cosine vx}. The $T$ and $v_x$ profiles change separately. The $T$ profile drops to the cold end at as early as $z \sim 0.2$ and hovers just above the cold phase temperature when the $v_x$ profile just starts to change at about the same point in space. In other words, this solution cools first (in position space), then mixes, meaning that there is significant change in $v_x$ within the cold phase. The pressure profile provides important insights into the physics that is driving the mixing layer. Naively, one would expect that the rapid cooling in the mixing layer leads to a region of low pressure at intermediate temperatures, which creates a pressure gradient that accelerates the inflow of hot gas. However, the top left panel of \autoref{fig:vx profile example sol} shows that this is clearly not the case. This suggests that it is turbulent mixing (which is in the form of effective turbulent viscosity and conductivity in our model), not pressure gradient, that controls the mass inflow rate and thus the radiative cooling rate. We further note that the hot phase pressure is slightly higher than that of the cold phase. This difference is accounted for by the additional ram pressure at the hot phase.

To better understand the physics that is controlling the behavior of the mixing layer, it is instructive to return to the fluid equations and deconstruct the advected enthalpy from the hot phase, or the temperature gradient, $\left. dT \right/ dz$, into its constituent terms. This can be done by solving for $\left. dT \right/ dz$ in \autoref{eq:T equation}. $\left. dT \right/ dz$ can be decomposed into terms that represent radiative cooling, conduction, viscous heating, and work done by the pressure gradient. By examining the relative importance of these terms, we can better understand the underlying physics that controls the mixing layer. 

The two bottom panels of \autoref{fig:vx profile example sol} shows the ``terms decomposition" in position ($z$) and  temperature space. There are three parts to the solution. Starting from the hot end, the mixing layer is initially dominated by conductive cooling. At intermediate temperatures, radiative cooling becomes significant and is balanced by conductive heating. This balance holds true until we reach the cold end, where the $x$-component of viscous heating takes over from conductive heating to balance the radiative cooling. This final part of the solution takes up a tiny fraction in temperature space but a significant portion of the position space. This is consistent with our previous observation that the $T$ and $v_x$ profiles change separately.

The top right panel of \autoref{fig:vx profile example sol} shows the column density and cooling flux distributions of the mixing layer (see \autoref{sec:phase distributions} for more details). Corresponding to the strong viscous heating zone near the cold end, we see a spike in the cooling flux distribution at low temperatures. This is a feature that is commonly seen in 3D simulations, and our analytic model is able to give us physical intuition towards the origin of this spike: radiating away the viscously heated material in the cold phase.

\subsubsection{Comparing Three TRMLs with different Choices of the Relative Shear Velocity $\mathcal{M}_{\rm rel}$}
\label{sec:compare three solutions}

Now that we are familiar with the key characteristics of a fiducial mixing layer, we expand our analysis to include three mixing layers with different choices of the relative shear velocity $\mathcal{M}_{\rm rel}$ (including a supersonic $\mathcal{M}_{\rm rel}$) and understand how the key characteristics change across these mixing layers.

\begin{figure*}
\centering
\includegraphics[width=\textwidth]{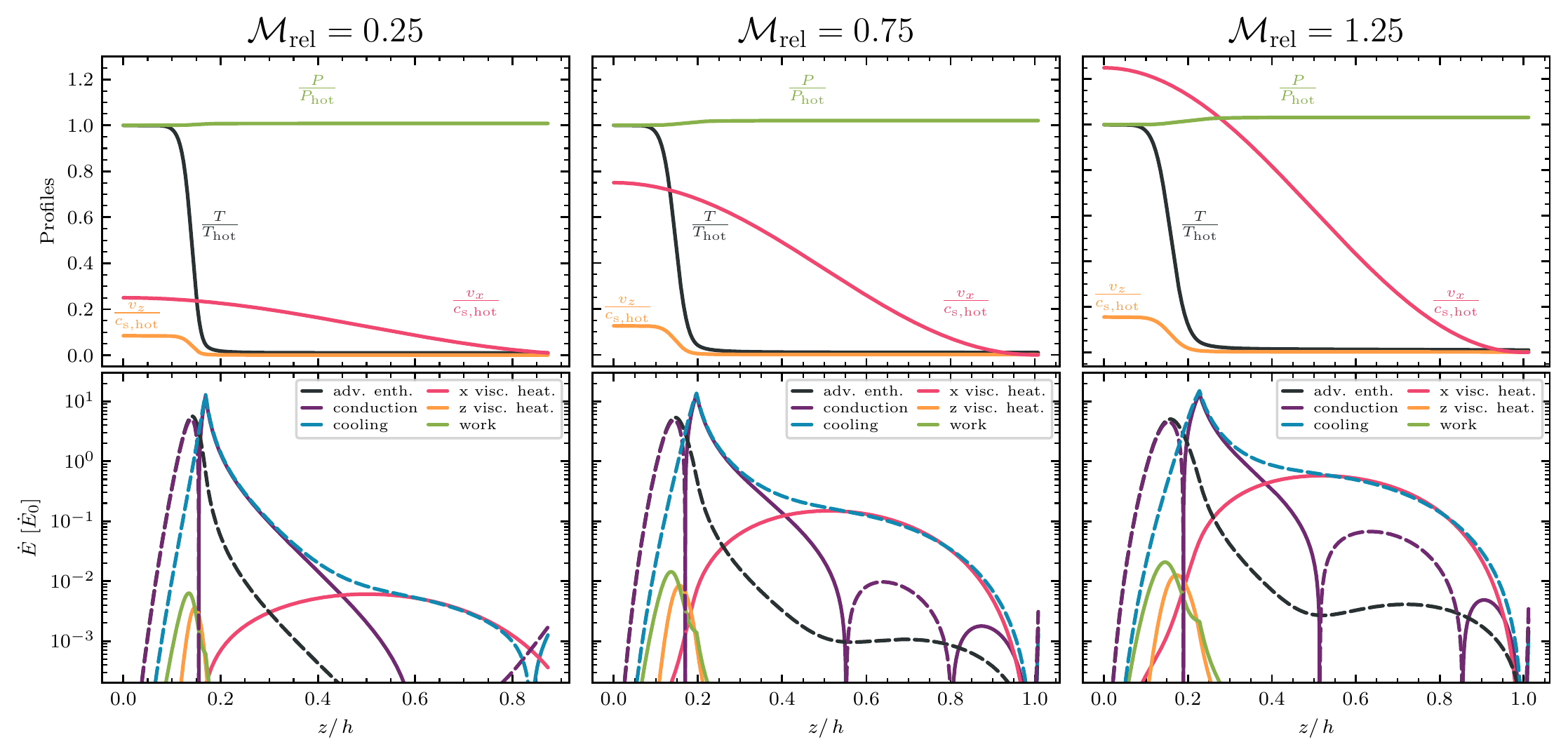}
\caption{The temperature, pressure, and velocity profiles (top) and the terms decomposition of \autoref{eq: before integration} (bottom) of solutions with the cosine $v_x$ profile at $\chi=10^2$, $\tau=10^{-1}$, $f_{\nu}=10^{-2}$, $\PR=0.1$ and $\mathcal{M}_{\rm rel}=0.25$, $0.75$, and $1.25$. The physical meaning of each of the curve is the same as in \autoref{fig:vx profile example sol}. Increasing $\mathcal{M}_{\rm rel}$ increases the magnitude of the $x$-component of viscous heating and make the transition to the region where viscous heating balances radiative cooling happen earlier. }
\label{fig:vx profile three vrels}
\end{figure*}

In \autoref{fig:vx profile three vrels}, we present the temperature, pressure, velocity profiles (top panels) and the terms decomposition of $\left. dT \right/ dz$ in position space (bottom panels) of three mixing layers with $\mathcal{M}_{\rm rel} =0.25$, $0.75$, and $1.25$. The most notable difference is that the magnitude of the $x$-component of viscous heating increases together with $\mathcal{M}_{\rm rel}$. This is expected because the $x$-component of viscous heating is proportional to $\left( \left. dv_x \right/ dz \right)^2$, and with a fixed thickness of the mixing layer and higher $\mathcal{M}_{\rm rel}$, $\left. dv_x \right/ dz$ is bound to increase in magnitude. On the other hand, most of the qualitative features of the mixing layer remain the same regardless of the choice of $\mathcal{M}_{\rm rel}$. All three mixing layers have a mixing zone where conduction and radiative cooling drives most of the temperature drop, and a viscous heating zone at the cold phase where the majority of the viscous heating takes place.

\begin{figure*}
\centering
\includegraphics[width=\textwidth]{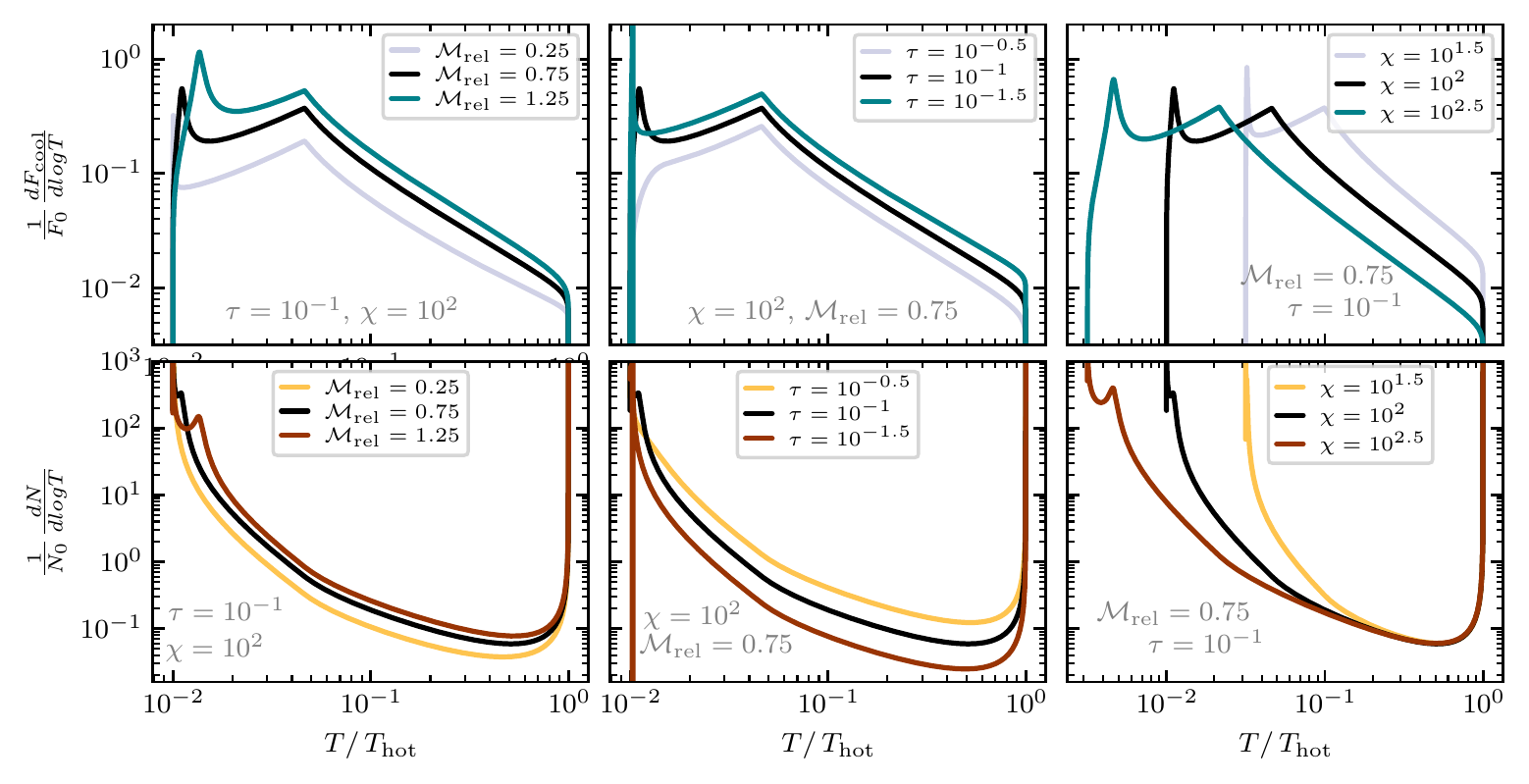}
\caption{The cooling flux (top) and column density (bottom) distributions of solutions with the cosine $v_x$ profile at $f_{\nu}=10^{-2}$, $\PR=0.1$ and different choices of $\mathcal{M}_{\rm rel}$, $\tau$, and $\chi$. The cooling flux distributions are normalized by
$F_0 = \rho_{\rm hot} c_{\rm s,hot}^3$, and the column density distributions are normalized by $N_0 = \left. \rho_{\rm hot} L_0 \right/ m_H$. In each column, the fiducial cooling flux and column density distributions, which are colored in black, has $\mathcal{M}_{\rm rel}=0.75$, $\chi=10^2$, and $\tau=10^{-1}$. Starting from this fiducial choice of parameters, we vary $\mathcal{M}_{\rm rel}$ (left), $\tau$ (middle), and $\chi$ (right) and compare the resulting cooling flux and column density distributions. In general, the power-law slopes of both distributions remain unaffected, but the height and shape of the spike at the cold end of the cooling flux distribution change as a function of all three parameters.}
\label{fig:phase distribution vs. vrel, t_cool_min, chi}
\end{figure*}

The left column of \autoref{fig:phase distribution vs. vrel, t_cool_min, chi} shows the cooling flux and column density distributions of the three mixing layers we just examined. As $\mathcal{M}_{\rm rel}$ increases, the spike at the cold phase of the cooling flux distribution increases in height, corresponding to the increased magnitude of the $x$-component of viscous heating. Furthermore, these spikes also gradually widen and shift towards higher temperature, indicating that the edges of the cooling and mixing zones in the mixing layers are starting to blend into each other as $\mathcal{M}_{\rm rel}$ increases. For the sake of completeness, we also show how the the cooling flux and column density distributions change as we vary $\tau$ and $\chi$ in the middle and right column of \autoref{fig:phase distribution vs. vrel, t_cool_min, chi}. In general, lower values of $\tau$ and $\chi$ correspond to a higher and narrower spike in the cooling flux distribution. 

As the mixing zone at the cold phase widens and undergoes more viscous heating with increasing $\mathcal{M}_{\rm rel}$, its mass content also increases, leading to the cold phase spikes in the column density distribution. Another interesting point to note is that the column density distributions clearly do not have the same shape as the $t_{\rm cool}$ curve as is predicted in pure cooling flows \citep[e.g.][]{Fabian:1994}. Above the temperature where the cooling curve peaks ($T_{\rm peak}$), the power law slope of the column density distribution remains negative and gradually levels off, but certainly does not turn positive, as does the $t_{\rm cool}$ curve.

\subsection{Bernoulli Flux and Requiring $\Hvisc= \left. \dot m v_{\rm rel}^2 \right/ 2$}\label{sec:Bernoulli flux}

In \autoref{sec:vx profile individual sol} we looked at the detailed structure of the mixing layers as a function of position and temperature by examining the relative importance of the terms in the fluid equations. Alternatively, it is also possible to treat the entire mixing layer as a whole. After deriving the equations we numerically solved at the beginning of \autoref{sec:methods}, we integrate from hot to cold to obtain a relationship between mass flux $\dot m$ and total cooling $\Qcool$ given by \autoref{eqn:sources and sink}. This is a statement of the energy sources and sinks in the entire mixing layer, and it helps us in understanding the energy budget of the system and allows us to explore how the importance of these energy sources and sinks change as parameters of our model are varied.

Besides the relationship between energy sources and sinks given by \autoref{eqn:sources and sink}, it is also crucial to enforce energy conservation on the mixing layer. The enthalpy flux (due to temperature difference between the hot and cold phase) and relative kinetic energy flux (due to shear velocity difference between the hot and cold phase) that is advected into the mixing layer must be radiated away by cooling to ensure energy conservation and prevent energy build-up within any part of the mixing layer. (We note that the kinetic energy influx must be first dissipated into heat, and then radiated away by cooling.) In other words, $\Qcool$ as defined in \autoref{eq:Qcool} must be balanced by what we define to be the ``Bernoulli flux", given by $ \dot m \left( \left. \Delta c_s^2 \right/ \left( \gamma-1 \right) + \left. v_{\rm rel}^2 \right/ 2 \right)$ \citep{Fielding:2022}. Another way to think about this is that the two phases are experiencing a perfectly inelastic collision through the mixing layer, and the relative kinetic energy between the phases must be dissipated into heat and radiated away in the mixing layer. Combining the Bernoulli flux equation with \autoref{eqn:sources and sink}, we demand that $\Hvisc + \mathcal{W} = \left. \dot m v_{\rm rel}^2 \right/ 2$. In practice, $\mathcal{W}$ is several orders of magnitude smaller than $\Hvisc$, which means our energy conservation condition simplifies to $\Hvisc = \left. \dot m v_{\rm rel}^2 \right/ 2$. 

Now that we have established why energy conservation dictates that $\Hvisc = \left. \dot m v_{\rm rel}^2 \right/ 2$ must be true for TRMLs, it is useful to examine what parameter choices for our analytic model satisfy this constraint. Given choices of $\mathcal{M}_{\rm rel}$, $\chi$, and $\tau$, which are the "physical" parameters that are set by the physics of the mixing layer of interest, it is crucial to realize that not all combinations of $\PR$ and $f_{\nu}$ produce solutions to TRML that satisfy $\Hvisc = \left. \dot m v_{\rm rel}^2 \right/ 2$. In other words, $\PR$ and $f_{\nu}$ are ``model-specific" parameters that are constrained by the energy conservation criterion, and the problem at hand is: how to pick the values for $\PR$ and $f_{\nu}$? In \autoref{sec:energy sources and sinks vs.Pr and fnu}, we will show that enforcing energy conservation reduces the two degrees of freedom in $\PR$ and $f_{\nu}$ by one. In \autoref{sec:matching mdot and Qcool}, we will compare our analytic model with 3D simulations produced by \cite{Fielding:2020} to clear up the remaining degree of freedom.

\subsection{Energy Sources \& Sinks vs. $\PR$ and $f_{\nu}$}
\label{sec:energy sources and sinks vs.Pr and fnu}

In this section, we keep all other parameters fixed at fiducial values and focus on how $\PR$ and $f_{\nu}$ affect the energy conservation condition by plotting the energy sources and sinks in \autoref{eqn:sources and sink} as we vary $\PR$ and $f_{\nu}$. 

\begin{figure*}
\centering
\includegraphics[width=\textwidth]{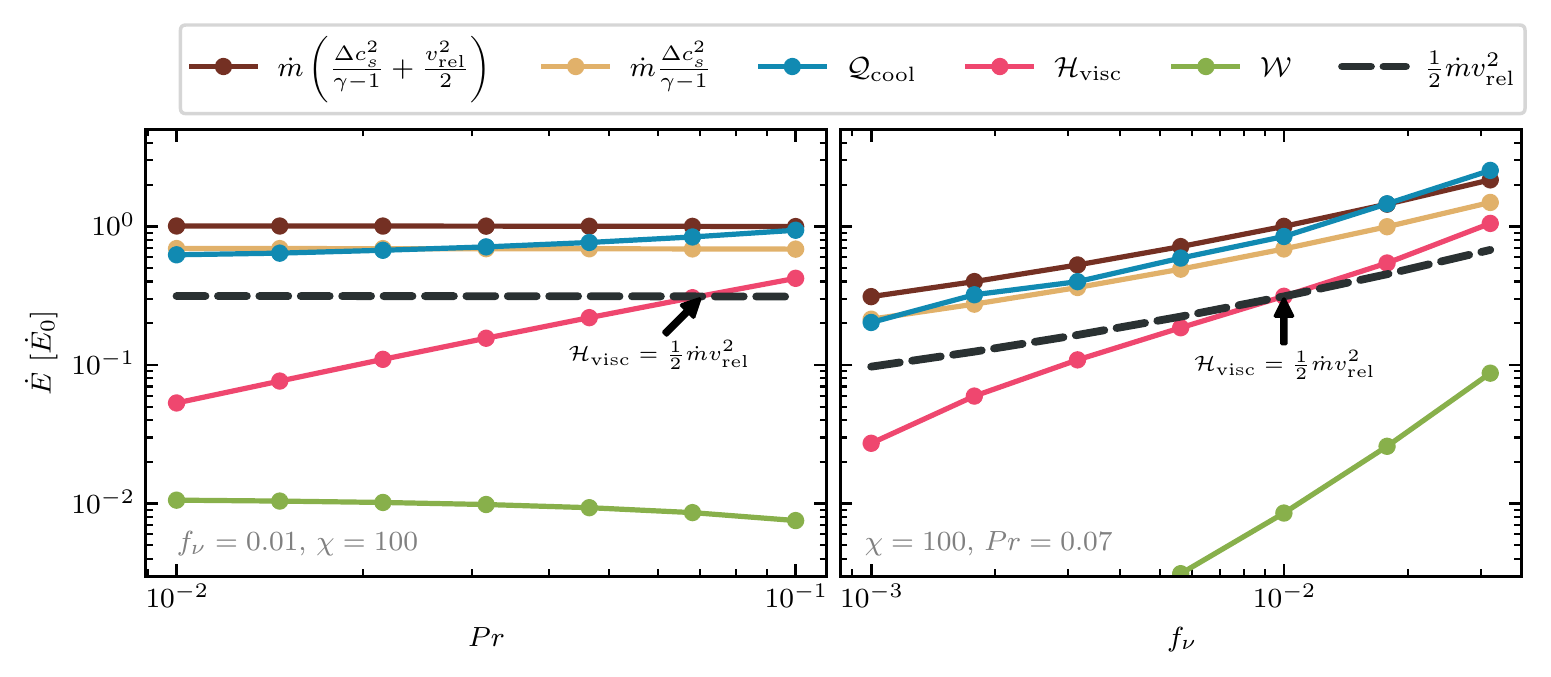}
\caption{The energy sources and sinks in \autoref{eqn:sources and sink} as a function of $\PR$ (left) and $f_{\nu}$ (right). As in \autoref{fig:vx profile example sol}, blue is related to radiative cooling, pink is related to viscous heating, and green is related to (work done by) pressure. Additionally, we show the Bernoulli flux (as defined in \autoref{sec:Bernoulli flux}) in dark brown, the enthalpy flux in light brown, and the relative kinetic energy flux in dotted black. Each vertical group of dots represents a solution to a mixing layer calculated using our analytic model with the cosine $v_x$ profile. All mixing layers have $\mathcal{M}_{\rm rel}=1.5$, $\tau=10^{-1}$, $\chi=10^2$, $\PR=0.07$, $f_{\nu}=10^{-2}$ unless either $\PR$ or $f_{\nu}$ is explicitly varied. In each panel, there is only one parameter choice, denoted by the black arrow, that satisfies the energy conservation condition $\Hvisc = \left. \dot m v_{\rm rel}^2 \right/ 2$ as discussed in \autoref{sec:Bernoulli flux}.}
\label{fig:energy sources and sinks vs. Pr, f_nu}
\end{figure*}

\autoref{fig:energy sources and sinks vs. Pr, f_nu} shows how the enthalpy flux, $\Qcool$, $\Hvisc$, and $\mathcal{W}$ in \autoref{eqn:sources and sink} and the Bernoulli flux defined in \autoref{sec:Bernoulli flux} vary with respect to $\PR$ and $f_{\nu}$. Each vertical group of dots represents a solution to a mixing layer calculated using our model. In each panel, the parameter choice that satisfies the energy conservation condition $\Hvisc = \left. \dot m v_{\rm rel}^2 \right/ 2$ can be found by identifying the intersection of $\Hvisc$ (the pink curve) and $\left. \dot m v_{\rm rel}^2 \right/ 2$ (the black dashed curve). For example, in the left panel we fix $f_{\nu}=0.01$ and vary $\PR$. $\Hvisc$ and $\left. \dot m v_{\rm rel}^2 \right/ 2$ intersects at around $\PR=0.07$. This is the only value of $\PR$ that yields a physically sensible solution for the given choice of $f_{\nu}$. A similar situation holds true in the right panels. Once we fix either $\PR$ or $f_{\nu}$, there is only one value of the other parameter that is consistent with energy conservation. In conclusion, demanding energy conservation ($\Hvisc = \left. \dot m v_{\rm rel}^2 \right/ 2$) reduces one degree of freedom in the parameter space of our analytic model, providing an important constraint on the choices of $\PR$ and $f_{\nu}$. We stress that this constraint does not come from simulations and is a natural result of enforcing energy conservation. In fact, in \autoref{sec:matching mdot and Qcool}, we will discuss how to use simulation results to further constrain the parameter space such that we can obtain best fit values of $\PR$ and $f_{\nu}$, the two parameters that are introduced by our analytic model.

\subsection{Energy Sources \& Sinks vs. $\mathcal{M}_{\rm rel}$ (with the Cosine $v_x$ Profile)}
\label{sec:vx profile, energy sources and sinks vs. vrel}

\begin{figure}
\centering
\includegraphics[width=\columnwidth]{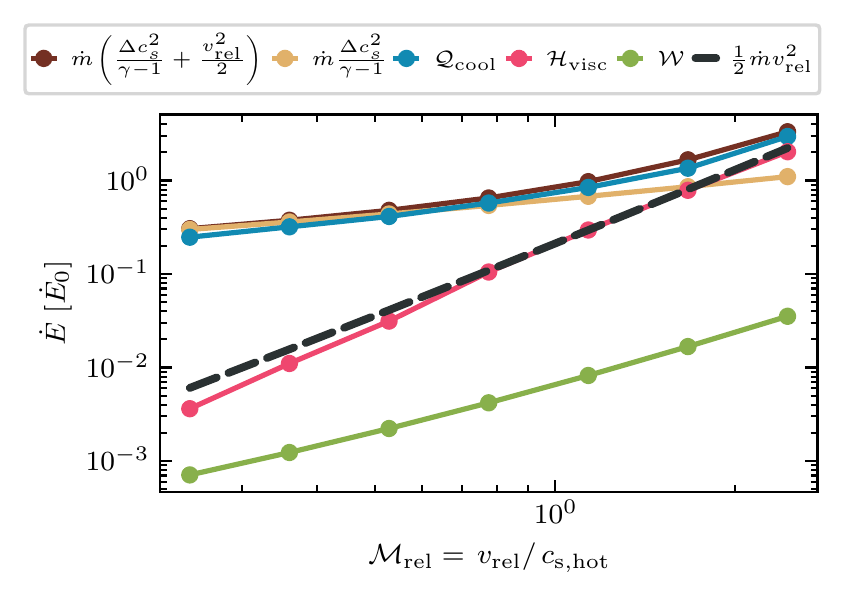}
\caption{The setup is similar to \autoref{fig:energy sources and sinks vs. Pr, f_nu}, but here we vary $\mathcal{M}_{\rm rel}$ for a series of mixing layers with the cosine $v_x$ profile, $\tau=10^{-1}$, $\chi=10^2$, $\PR=0.07$, and $f_{\nu}=10^{-2}$. Increasing $\mathcal{M}_{\rm rel}$ implies increasing $\Hvisc$, as shown by the upward sloping $\Hvisc$ curve. As we saw in \autoref{fig:vx profile example sol}, most of the contribution to $\Hvisc$ comes from the region near the cold phase where viscous heating balances radiative cooling. Thus, as $\Hvisc$ (pink) continues to increase with $\mathcal{M}_{\rm rel}$ and becomes comparable to the enthalpy flux, we see an upturn in $\Qcool$ (blue) as a response.}
\label{fig:vx profile, energy sources and sinks vs. vrel}
\end{figure}

Our controlled experiments so far in which we varying $\PR$ and $f_{\nu}$ are mainly for providing constraints on the viable choices of these parameters. Alternatively, it is also instructive to vary the other parameters of our analytic model that are set by the physics of the mixing layer of interest. Physically, this corresponds to comparing the energy budget across an ensemble of mixing layers and can potentially help us gain insights into what is controlling the phase structure. In this section, we explore the consequences of vary $\mathcal{M}_{\rm rel}$ on the energy sources and sinks in a mixing layer.

As shown in \autoref{fig:vx profile, energy sources and sinks vs. vrel}, with $\tau=10^{-1}$, $\chi=10^2$, $\PR=0.07$, and $f_{\nu}=10^{-2}$, energy conservation is satisfied (i.e., $ \dot m \left( \left. \Delta c_s^2 \right/ \left( \gamma-1 \right) + \left. v_{\rm rel}^2 \right/ 2 \right) = \Qcool + \Hvisc + \mathcal{W}$) for a range of $\mathcal{M}_{\rm rel}$. The crucial point to note is that the choice of $\mathcal{M}_{\rm rel}$ affects the energy conservation criterion very weakly. Graphically, this is shown by the overlap of the curves representing $\Hvisc$ and $\left. \dot m v_{\rm rel}^2 \right/ 2$ in the entire range of $\mathcal{M}_{\rm rel}$ plotted in \autoref{fig:vx profile, energy sources and sinks vs. vrel}. This characteristic ensures that the ensemble of mixing layers at different $\mathcal{M}_{\rm rel}$ (with all other parameters held fixed) all satisfy energy conservation, which makes comparing between them physically meaningful. 

As discussed in \autoref{sec:vx profile individual sol}, higher $\mathcal{M}_{\rm rel}$ implies more viscous heating ($\Hvisc$). This can be seen in \autoref{fig:vx profile, energy sources and sinks vs. vrel} through the positively sloped $\Hvisc$ line (in pink). While analyzing the detailed structure of these mixing layers with the cosine $v_x$ profile prescription in \autoref{sec:fiducial vrel solution}, we observed that most of the viscous heating is concentrated in the third (``viscous") zone within the cold phase, where viscous heating is balanced by radiative cooling. Thus, as $\Hvisc$ increases with $\mathcal{M}_{\rm rel}$, $\Qcool$ increases together with it. This effect is most pronounced at the high $\mathcal{M}_{\rm rel}$ end of \autoref{fig:vx profile, energy sources and sinks vs. vrel}, where the curves for the enthalpy flux and the Bernoulli flux (see \autoref{sec:Bernoulli flux}) diverge and $\Qcool$ follows the Bernoulli flux. The importance of including the relative kinetic energy flux in our energy conservation argument is most apparent in the high $\mathcal{M}_{\rm rel}$ regime in which $\Hvisc$ can exceed the enthalpy flux ($\dot m \Delta c_s^2 / \left( \gamma-1 \right)$).

The essential point is that at low $\Mach$ the enthalpy flux far exceed the viscous heating, but the enthalpy flux scales with $\Machrel^{3/4}$, as is found in 3D simulations \citep{Fielding:2020,Tan:2021}, whereas the viscous heating scales with $\Machrel^2$, so at sufficiently large $\Machrel$ the viscous heating will always dominate. When the viscous heating begins to dominate the $\Qcool$, which had been primarily following the enthalpy flux scaling, now must steepen to follow the viscous heating scaling. This upturn in $\Qcool$ is an important feature of the 3D simulations of the mixing layers (this point will be further explored in \autoref{sec:vx, momentum, 3D, energy sources and sinks vs. vrel}). We have seen that this feature, as well as the cold end spike in the cooling flux distribution discussed in \autoref{sec:fiducial vrel solution}, are both produced by the extended viscous zone at the cold phase of the mixing layer. We believe the same basic physics applies to the 3D simulations because they exhibit very similar features.

Although \autoref{fig:vx profile, energy sources and sinks vs. vrel} does provide many useful insights, it does not reproduce all of the key features that are seen in the 3D simulations. Namely, in the 3D simulations we also see a downturn in the enthalpy flux with increasing $\mathcal{M}_{\rm rel}$ (we will later show this explicitly in \autoref{fig:vx, momentum, 3D, energy sources and sinks vs. vrel}), which is not a feature of our analytic model with the cosine $v_x$ profile setup. We therefore turn to the alternate version of our analytic model in which we prescribe the $x$-momentum $\rho v_x$ as a cosine and infer the value of $\vx$ from that. The detailed setup of this model is introduced in \autoref{sec:momentum profile setup}. Now we explore key differences of the mixing layer structure we obtained from the $\vx$ and $\rho v_x$ profiles to see what the $\rho v_x$ profile setup can tell us about the underlying physics of the 3D simulations.

\subsection{Energy Sources \& Sinks vs. $\mathcal{M}_{\rm rel}$ (with the Cosine $\rho v_x$ Profile)}
\label{sec:momentum profile, energy sources and sinks vs. vrel}

To better understand how the energy sources and sinks vary with $\mathcal{M}_{\rm rel}$ under the cosine $\rho v_x$ profile prescription, it is instructive to first go back and examine the structure of a mixing layer under this new setup.

\begin{figure*}
\centering
\includegraphics[width=\textwidth]{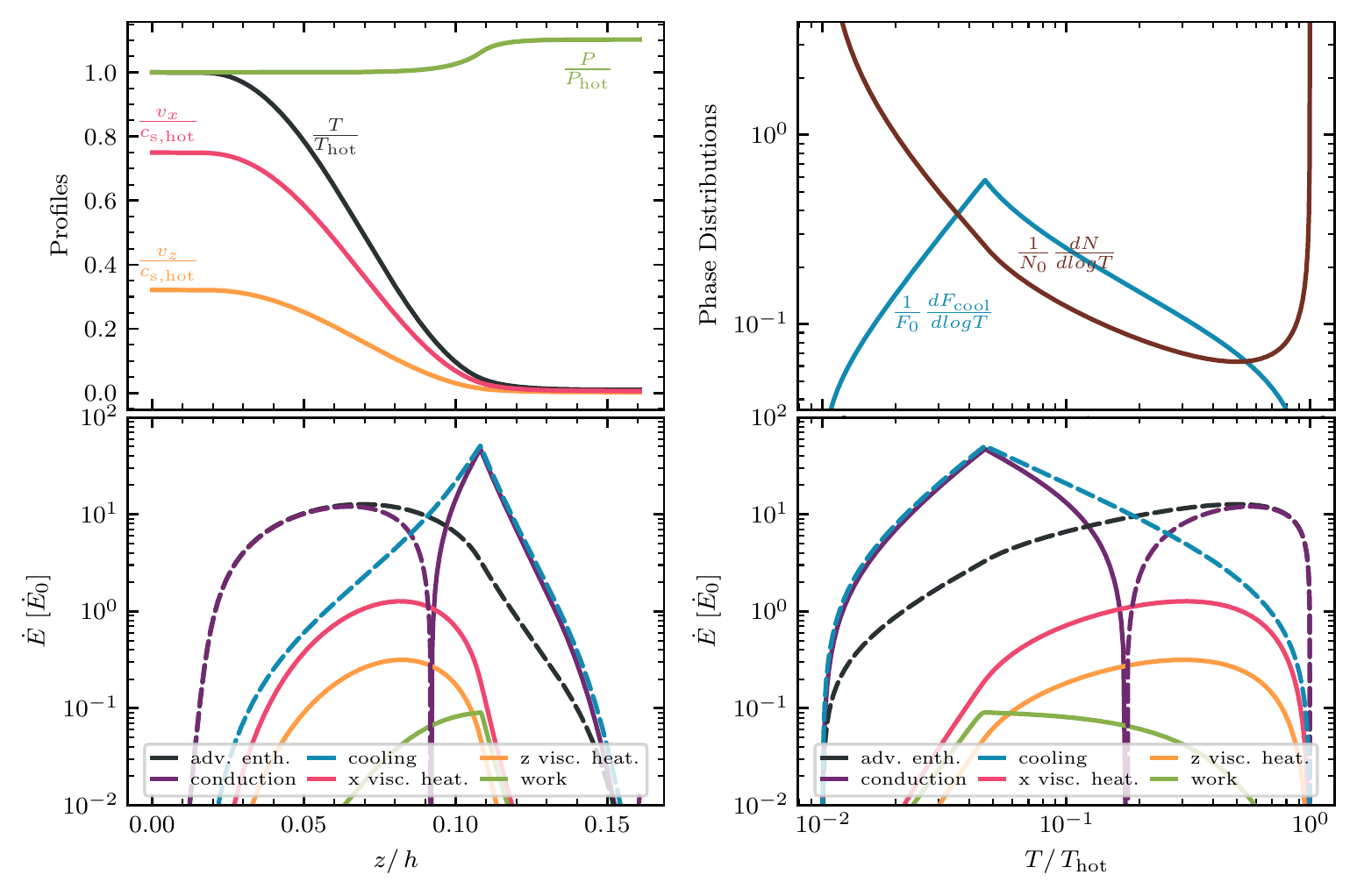}
\caption{A fiducial solution of our analytic model using the cosine $\rho v_x$ profile. The parameter choices for this solution are: $\mathcal{M}_{\rm rel}=0.75$, $\chi=10^2$, $\tau=10^{-1.5}$, $f_{\nu}=10^{-3}$, $\PR=0.3$. The information in each panel and the color choices is the same as \autoref{fig:vx profile example sol}. However, a crucial difference is that mixing layers with the cosine $\rho v_x$ profile has the temperature and velocity profiles change simultaneously, which means there is no significant viscous heating near the cold phase and thus no spike in the cooling flux distribution. Now viscous heating is mostly concentrated at intermediate temperatures and never dominant amongst the various contributions to the advected enthalpy.}
\label{fig:momentum profile example sol}
\end{figure*}

\autoref{fig:momentum profile example sol} shows a mixing layer with the cosine $x$-momentum profile and fiducial choice of parameters. Compared to a mixing layer with the cosine $v_x$ profile (\autoref{fig:vx profile example sol}), the crucial difference here is that the $T$ and $v_x$ profiles change almost simultaneously in position space. This is because we are tying $\left. d \rho \right/ dz$ and $\left. d v_x \right/ dz$ together through the cosine $\rho v_x$ profile, and since $\left. d \rho \right/ dz$ and $\left. d T \right/ dz$ are inherently tied through the ideal gas law, $\left. d v_x \right/ dz$ and $\left. d T \right/ dz$ are also related, leading to the simultaneous change of $T$ and $v_x$. A direct consequence is that we lose the extended viscous zone at the cold phase and the corresponding spike in the cooling flux distribution. Instead, both the $x$- and $z$-component of viscous heating are now strongest at intermediate temperatures in the mixing layer.

\begin{figure}
\centering
\includegraphics[width=\columnwidth]{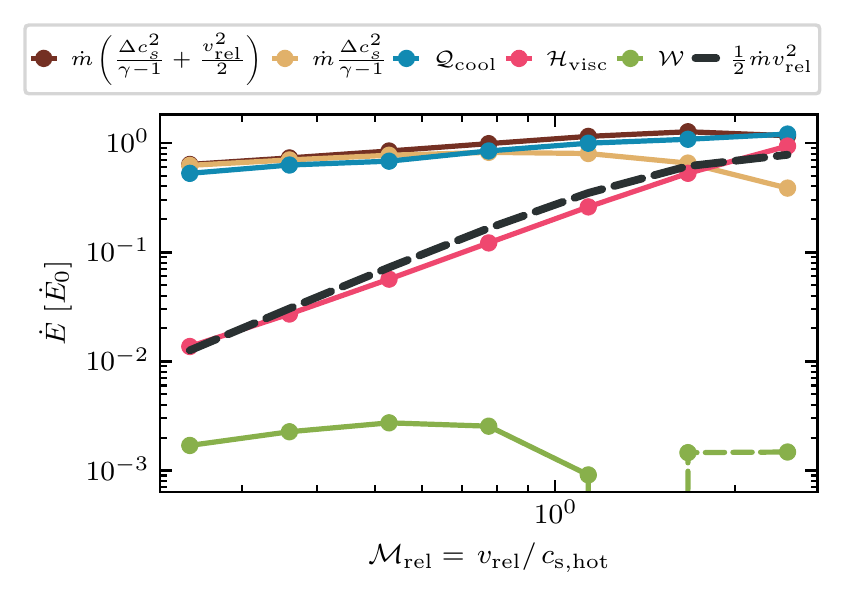}
\caption{Similar to \autoref{fig:vx profile, energy sources and sinks vs. vrel} but with the cosine $\rho v_x$ profile, $\tau=10^{-1.5}$, $\PR=0.3$, $\chi=10^2$, and $f_{\nu}=10^{-3}$. As shown in \autoref{fig:momentum profile example sol}, with the cosine $\rho v_x$ profile we lose the region where viscous heating balances radiative cooling, which means increasing $\Hvisc$ with $\mathcal{M}_{\rm rel}$ does not affect $\Qcool$. To maintain energy conservation (energy radiated away=enthalpy flux from hot phase+viscous heating), there is a downturn in enthalpy flux to account for the enhanced viscous heating.}
\label{fig:momentum profile, energy sources and sinks vs. vrel}
\end{figure}

Recall that our energy conservation argument introduced in \autoref{sec:Bernoulli flux} states that energy radiated away through cooling is balanced by the enthalpy flux plus the viscous dissipation. As we turn up viscous dissipation by increasing $\mathcal{M}_{\rm rel}$, as shown in \autoref{fig:momentum profile, energy sources and sinks vs. vrel}, the amount of radiative cooling is unaffected because at intermediate temperatures, where viscous heating is the strongest, radiative cooling is significantly larger in magnitude. Thus, in order to maintain energy conservation, the enthalpy flux decreases with increasing $\mathcal{M}_{\rm rel}$ to compensate for the increase in viscous dissipation. Graphically, this is manifested by the downturn in the brown curve for enthalpy flux in \autoref{fig:momentum profile, energy sources and sinks vs. vrel}.

We note that this downturn in enthalpy flux is another important feature of the 3D simulations. This investigation with the $x$-momentum profile allows us to link this feature with significant viscous heating at intermediate temperatures.

\subsection{Finding $\PR$ and $f_{\nu}$ Values that Best Match $\dot m$ and $\Qcool$ from 3D Simulations}
\label{sec:matching mdot and Qcool}

We have argued that our analytic model with both the cosine $v_x$ and the cosine $\rho v_x$ profiles are able to reproduce key features of the 3D simulations (upturn in $\Qcool$ discussed in \autoref{sec:vx profile, energy sources and sinks vs. vrel} and downturn in enthalpy flux discussed in \autoref{sec:momentum profile, energy sources and sinks vs. vrel}). To provide concrete evidence for this argument, we seek to directly compare our analytic model with the 3D simulations. To do so, we need to return to the question of finding appropriate values for $\PR$ and $f_{\nu}$, the two additional parameters introduced by our analytic model.

Remember in \autoref{sec:energy sources and sinks vs.Pr and fnu} we have already discussed how we can reduce one degree of freedom in the parameter space of $\PR$ and $f_{\nu}$ by enforcing energy conservation. Here we constrain the remaining degree of freedom by requiring that our analytic solution with some $\mathcal{M}_{\rm rel}$, $\chi$, and $tau$ reproduces key results of the 3D simulation with the same parameter choices. We are most interested in the mass, momentum, and energy transfer between the two phases and the shape of the phase distributions. Thus, it is the most useful to compare the values of the mass flux $\dot m$ and the total amount of cooling in the mixing layer $\Qcool$. In particular, we seek an analytic model that produces $\dot m$ and $\Qcool$ that are identical to its 3D simulation counterpart with the same set of physical parameters ($\mathcal{M}_{\rm rel}$, $\chi$, and $\tau$).

Suppose we were given a 3D simulation with some relative shear velocity $\mathcal{M}_{\rm rel}$, characteristic cooling time $\tau$, density contrast $\chi$, and we select a time in the simulation such that we have a well-defined value for $h$, the thickness of the mixing layer. Then we can plug the values of these parameters into our analytic model to obtain the corresponding 1.5D solutions. Note that since there is still one degree of freedom in the $\PR$ vs. $f_{\nu}$ parameters, there is a suite of 1.5D solutions (with various combinations of $\PR$ and $f_{\nu}$) that satisfy energy conservation and can potentially match the 3D simulation we started with. Each of these 1.5D solutions, or each ``energy-conserving" pair of $\PR$ and $f_{\nu}$, yields a mixing layer with some values of $\dot m$ and $\Qcool$. Together this suite of mixing layers form a linear subset in the $\dot m$--$\Qcool$ space. To determine which pair of $\PR$ and $f_{\nu}$ best matches the 3D simulation and how good the match is, we simply take the values of $\dot m$ and $\Qcool$ from the 3D simulation as a point in the $\dot m$--$\Qcool$ space and plot it together with linear subset in the $\dot m$--$\Qcool$ space formed by the suite of mixing layers from our analytic model. The point on the linear subset that is the closest to the 3D simulation point corresponds to the best fit $\PR$ and $f_{\nu}$ values. 

\begin{figure}
\centering
\includegraphics[width=\columnwidth]{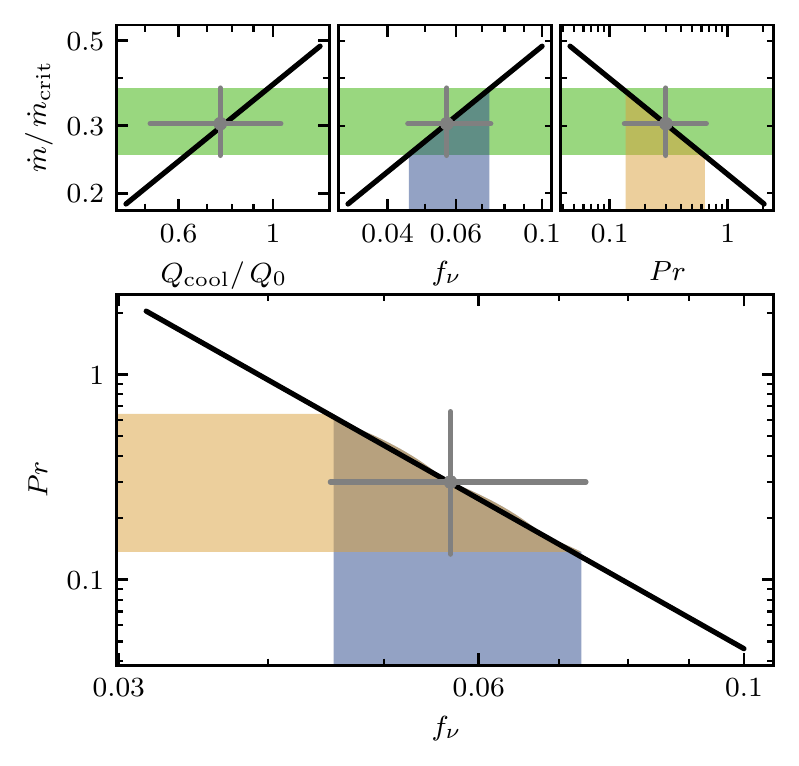}
\caption{Schematic of finding the best fit $\PR$ and $f_{\nu}$ for a 3D simulation. Consider a 3D simulation with $\mathcal{M}_{\rm rel}=0.75$, $\tau=10^{-1.5}$, and $\chi=10^2$. We seek to find a solution using our analytic model with the cosine $v_x$ profile that best matches the 3D simulation. The complication is that our analytic model takes in two additional parameters, $\PR$ and $f_{\nu}$, as inputs. Imposing the energy conservation condition $\Hvisc = \left. \dot m v_{\rm rel}^2 \right/ 2$ reduces one degree of freedom in the parameter space. The energy conserving pairs of $\PR$ and $f_{\nu}$ are plotted in the bottom panel. Each pair of $\PR$ and $f_{\nu}$ produces an analytic solution with its own mass flux ($\dot m$) and $\Qcool$. The corresponding $\dot m$ vs. $\Qcool$, $f_{\nu}$, and $\PR$ plots are shown in the top row. In the top left panel, $\dot m$ is normalized by $\dot m_{\rm crit} = \rho_{\rm hot} c_{\rm s,hot}$, and $\Qcool$ is normalized by $Q_0 = \rho_{\rm hot} c_{\rm s,hot}^3 L_0$. The grey point (with error bars) on the top left panel shows the 3D simulation values of $\dot m$ and $\Qcool$. Note that the 3D simulation point lies exactly on the curve generated from our analytic model, meaning that there is a combination of $\PR$ and $f_{\nu}$ that, when plugged into our analytic model, is able to accurately reproduce both $\dot m$ and $\Qcool$ from the 3D simulation. The green shaded region denotes the 3D simulation value of $\dot m$ with $2\sigma$ uncertainty, which can be extended to the $\dot m$ vs. $f_{\nu}$ and $\PR$ plots to find the corresponding best fit values of $f_{\nu}$ and $\PR$ with $2\sigma$ uncertainty, which are the blue and orange shaded regions, respectively. The blue and orange shaded regions are also plotted together in the bottom panel.}
\label{fig:schematic of best fit Pr and f_nu}
\end{figure}

\autoref{fig:schematic of best fit Pr and f_nu} is a graphical representation of this idea. We have chosen the cosine $v_x$ profile setup and fiducial choices of parameters ($\mathcal{M}_{\rm rel}=0.75$, $\tau=10^{-1.5}$, $\chi=10^2$) for this demonstration. For any given value of $f_{\nu}$, the corresponding value of $\PR$ that satisfies the energy conservation condition $\Hvisc = \left. \dot m v_{\rm rel}^2 \right/ 2$ can be found by performing a bisection on $\PR$. We are able to perform this bisection because, as shown in the left panel of \autoref{fig:energy sources and sinks vs. Pr, f_nu}, when $\PR$ is greater than the energy-conserving value, $\Hvisc > \left. \dot m v_{\rm rel}^2 \right/ 2$, and when $\PR$ is smaller than the energy-conserving value, $\Hvisc < \left. \dot m v_{\rm rel}^2 \right/ 2$. (Notice on the right panel of \autoref{fig:energy sources and sinks vs. Pr, f_nu} that the same holds true in the $f_{\nu}$ parameter space, which means this bisection can also be perform on $f_{\nu}$ for a given value of $\PR$.) Note that this bisection on $\PR$ is different from the bisection on the mass flux $\dot m$ as discussed in \autoref{sec:bisection on mdot}. For each ``guess" of $\PR$, we first bisect on $\dot m$ to find a temperature profile that reaches the cold phase smoothly. Then we compare $\Hvisc$ and $\left. \dot m v_{\rm rel}^2 \right/ 2$ for the mixing layer with the eigenvalue of $\dot m$ to inform our next guess of $\PR$. We repeat this bisection on $\PR$ for many values of $f_{\nu}$ to obtain a suite of energy-conserving pairs of $\PR$ and $f_{\nu}$. Together these pairs form a curve on the bottom panel of \autoref{fig:schematic of best fit Pr and f_nu}.

Each energy-conserving pair of $\PR$ and $f_{\nu}$ corresponds to a mixing layer with some $\dot m$ and $\Qcool$. Thus, for each point on the $\PR$--$f_{\nu}$ curve on the bottom panel of \autoref{fig:schematic of best fit Pr and f_nu}, we can find corresponding values of $\dot m$ and $\Qcool$, allowing us to make plots of $\dot m$ versus $\Qcool$, $f_{\nu}$, and $\PR$, as shown in the top panels of \autoref{fig:schematic of best fit Pr and f_nu}. 

We compare to a 3D simulation with the exact same parameter choices and include the result as a point in the $\dot m$--$\Qcool$ space, with error bars denoting the $2\sigma$ temporal variations around when $h$ in the 3D simulation is within 5 \% of $h=1$. The 3D simulation point appears to lie almost exactly on the $\dot m$--$\Qcool$ curve from our analytic model (top left panel of \autoref{fig:schematic of best fit Pr and f_nu}), which means there exists some energy-conserving pair of $\PR$ and $f_{\nu}$ that, when plugged into our analytic model, very closely reproduces both $\dot m$ and $\Qcool$ of the 3D simulation. The best-fit values of $\PR$ and $f_{\nu}$ (with error bars) can be found by projected the $\dot m$ value from the 3D simulation onto the $\dot m$ versus $f_{\nu}$ and $\PR$ spaces, as shown in the top middle and top right panels of \autoref{fig:schematic of best fit Pr and f_nu}. The inferred values of $\PR$ and $f_{\nu}$ are also shown on the $\PR$ versus $f_{\nu}$ curve in the bottom panel. 

\begin{figure*}
\centering
\includegraphics[width=\textwidth]{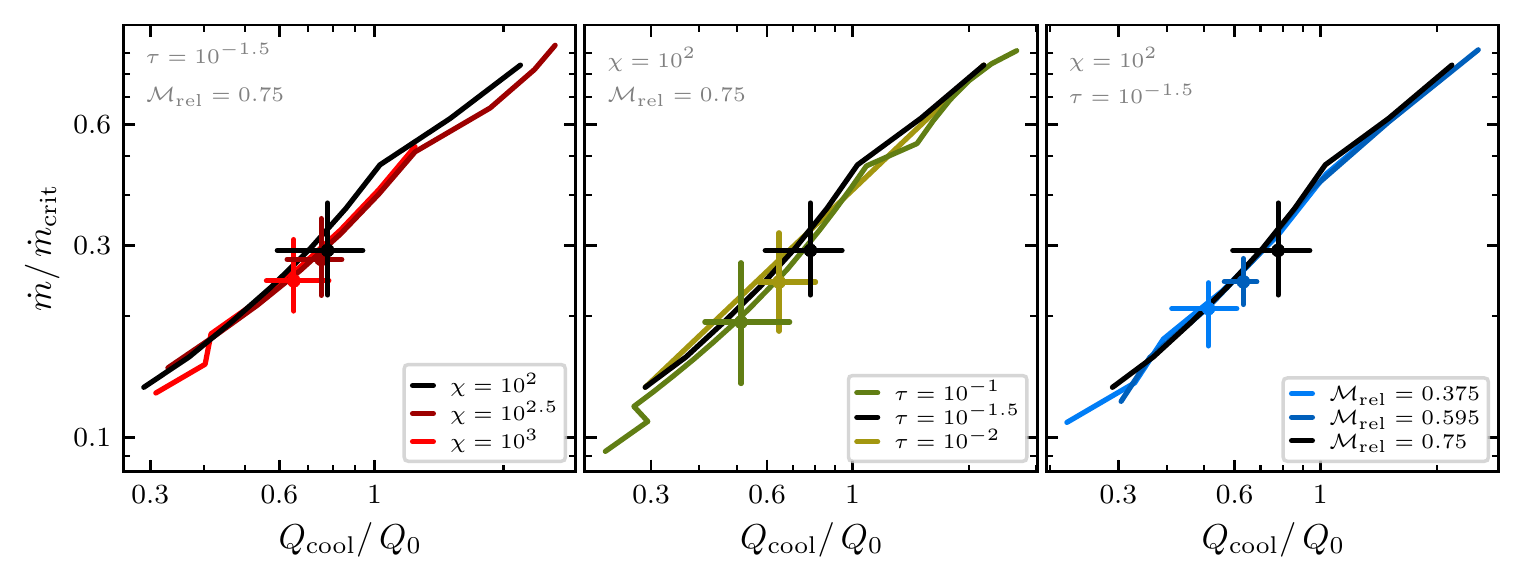}
\caption{The $\dot m$ vs. $\Qcool$ curves for solutions with the cosine $v_x$ profile at different choices of $\mathcal{M}_{\rm rel}$, $\tau$, and $\chi$. As described in \autoref{fig:schematic of best fit Pr and f_nu}, the $\dot m$ vs. $\Qcool$ curves are obtained from the $\PR$ vs. $f_{\nu}$ curve that satisfies energy conservation. $\dot m$ is normalized by $\dot m_{\rm crit} = \rho_{\rm hot} c_{\rm s,hot}$, and $\Qcool$ is normalized by $Q_0 = \rho_{\rm hot} c_{\rm s,hot}^3 L_0$. In each panel, the $\dot m$ vs. $\Qcool$ curve with the fiducial choice of parameters, $\mathcal{M}_{\rm rel}=0.75$, $\tau=10^{-1.5}$, and $\chi=10^2$, are colored in black. We vary $\chi$ (left), $\tau$ (middle), and $\mathcal{M}_{\rm rel}$ (right) and compare the resulting $\dot m$ vs. $\Qcool$ curves. The $\dot m$ and $\Qcool$ combinations of the corresponding 3D simulations are also plotted. The 3D simulation points all lie on the curves from our analytic model, which means there is a combination of $\PR$ and $f_{\nu}$ that, when plugged into our analytic model, is able to accurately reproduce both $\dot m$ and $\Qcool$ from the 3D simulation. This also means that the 3D simulations satisfies the energy conservation condition $\Qcool = \left( \left. \Delta c_s^2 \right/ \left( \gamma-1 \right) + \left. v_{\rm rel}^2 \right/ 2 \right) \dot m$, which is imposed to be true along the curves from our analytic model.}
\label{fig:vx profile, match mdot and Q_cool}
\end{figure*}

\begin{figure*}
\centering
\includegraphics[width=\textwidth]{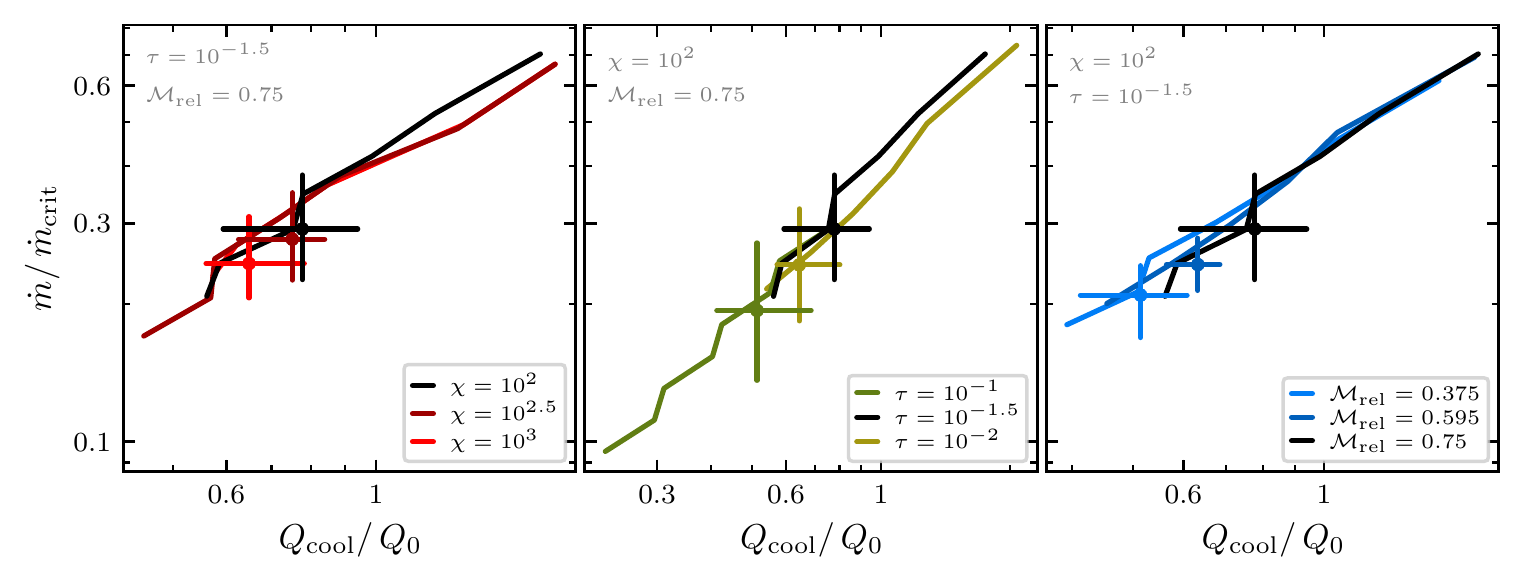}
\caption{Same as \autoref{fig:vx profile, match mdot and Q_cool}, but with the cosine $\rho v_x$ profile.}
\label{fig:momentum profile, match mdot and Q_cool}
\end{figure*}

\begin{figure*}
\centering
\includegraphics[width=\textwidth]{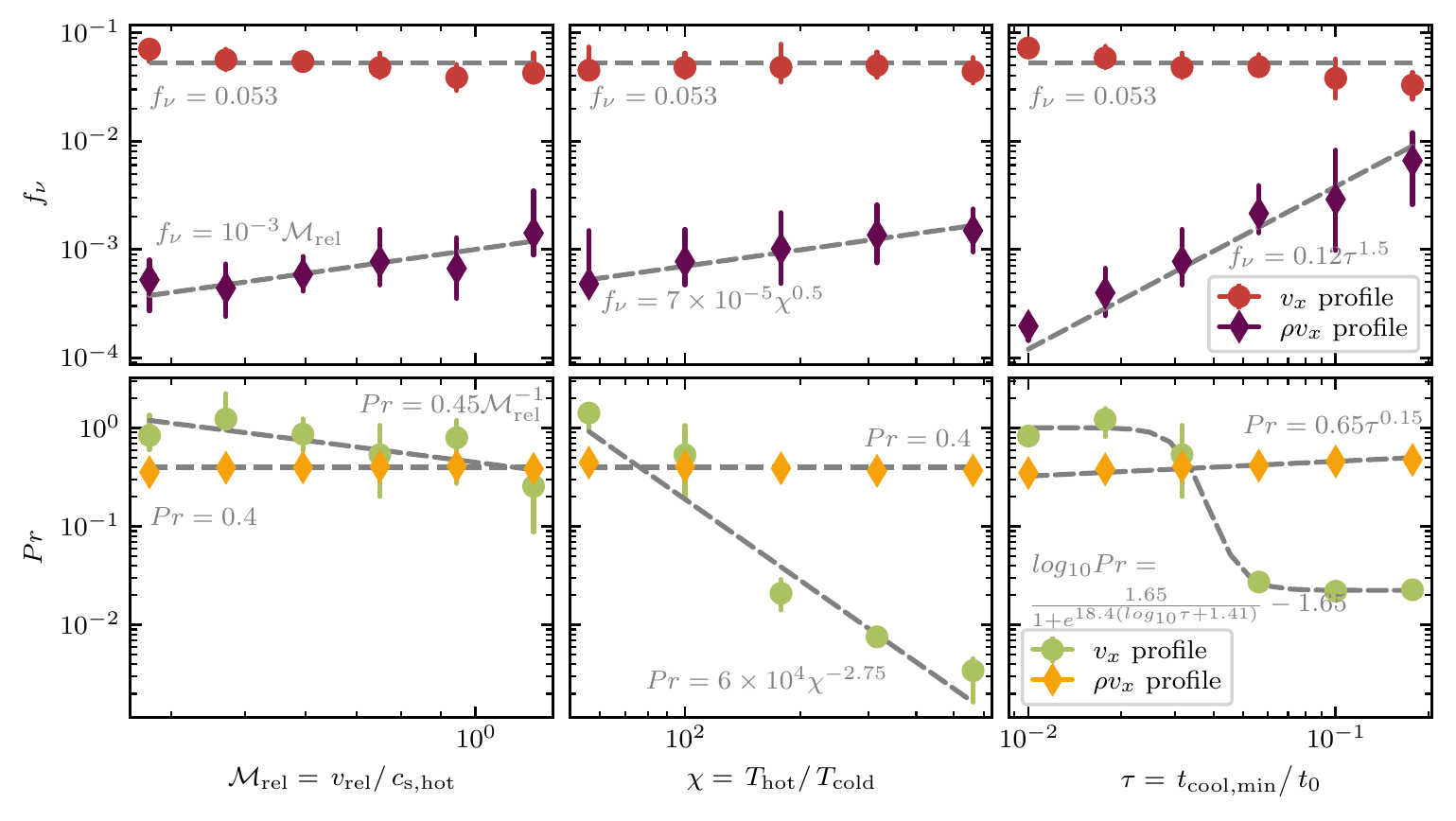}
\caption{Best fit $\PR$ and $f_{\nu}$  as a function of $\mathcal{M}_{\rm rel}$, $\chi$, and $\tau$ for the cosine $v_x$ and $\rho v_x$ profiles. The fiducial choice of parameters are $\mathcal{M}_{\rm rel}=0.75$, $\tau=10^{-1.5}$, and $\chi=10^2$, and we vary $\mathcal{M}_{\rm rel}$, $\chi$, and $\tau$ in the left, middle and right columns. Each value of $\PR$ and $f_{\nu}$ is obtained by comparing with the corresponding 3D simulation and performing the procedure described by \autoref{fig:schematic of best fit Pr and f_nu}. The grey dotted lines indicate how $\PR$ and $f_{\nu}$ scale with $\mathcal{M}_{\rm rel}$, $\chi$, and $\tau$. These scaling relationships can be used to directly calculate the best fit $\PR$ and $f_{\nu}$ for any combination of $\mathcal{M}_{\rm rel}$, $\chi$, and $\tau$ (without resorting to the computationally expensive bisection in the $\PR$ vs. $f_{\nu}$ parameter space).}
\label{fig:best fit Pr and f_nu for the vx and momentum profile}
\end{figure*}

In \autoref{fig:vx profile, match mdot and Q_cool} and \autoref{fig:momentum profile, match mdot and Q_cool}, we repeat the same exercise for several choices of $\mathcal{M}_{\rm rel}$, $\chi$, and $\tau$, with both the $v_x$ and the $x$-momentum profile. We see that in all cases the 3D simulation points lie very close to the curves obtained from our analytic model, which means both the $v_x$ and $\rho v_x$ profiles are able to reproduce $\dot m$ and $\Qcool$ values from the 3D simulations. The analytic model also has the additional benefit of helping us determine the turbulent Prandtl number, which is information not immediately available from the 3D simulations. The best fit $\PR$ and $f_{\nu}$ values for each choice of parameters can be determined using the approach illustrated in \autoref{fig:schematic of best fit Pr and f_nu} and is shown in \autoref{fig:best fit Pr and f_nu for the vx and momentum profile} as a function of $\mathcal{M}_{\rm rel}$, $\chi$, and $\tau$. For the cosine $v_x$ profile setup, the best fit $f_{\nu}$ values stay constant across different $\mathcal{M}_{\rm rel}$, $\chi$, and $\tau$, and the best fit $\PR$ values can be closely approximated by power-laws with respect to $\mathcal{M}_{\rm rel}$ and $\chi$, and a logistic function in log space with respect to $\tau$ . As for the cosine $\rho v_x$ profile, $\PR$ stays almost constant (there is a very weak dependence on $\tau$) and $f_{\nu}$ shows power-law variations with respect to all three parameters.

To summarize results from \autoref{fig:best fit Pr and f_nu for the vx and momentum profile}, we provide equations that allow us to calculate the best fit values of $\PR$ and $f_{\nu}$ as a function of $\mathcal{M}_{\rm rel}$, $\chi$, and $\tau$. For the cosine $v_x$ profile, the best fit values of $\PR$ and $f_{\nu}$ are given by

\begin{align}
    &\PR = \left( \frac{\mathcal{M}_{\rm rel}}{0.75} \right)^{-1} \left( \frac{\chi}{10^2} \right)^{-2.75} 10^{\frac{1.65}{1 + e^{18.4({\rm log}_{10} \tau + 1.41)}}},\\
    &f_{\nu} = 0.053.
\end{align}

For the cosine $\rho v_x$ profile, the best fit values of $\PR$ and $f_{\nu}$ are given by

\begin{align}
    &\PR = 0.4 \left( \frac{\tau}{10^{-1.5}} \right)^{0.15},\\
    &f_{\nu} = 0.00065 \left( \frac{\mathcal{M}_{\rm rel}}{0.75} \right) \left( \frac{\chi}{10^2} \right)^{-0.5} \left( \frac{\tau}{10^{-1.5}} \right)^{1.5}.
\end{align}

\subsection{Comparing Energy Sources \& Sinks vs. $\mathcal{M}_{\rm rel}$, $\chi$, and $\tau$ from the Two Variations of the Analytic Model} and 3D Simulations
\label{sec:vx, momentum, 3D, energy sources and sinks vs. vrel}

\begin{figure*}
\centering
\includegraphics[width=\textwidth]{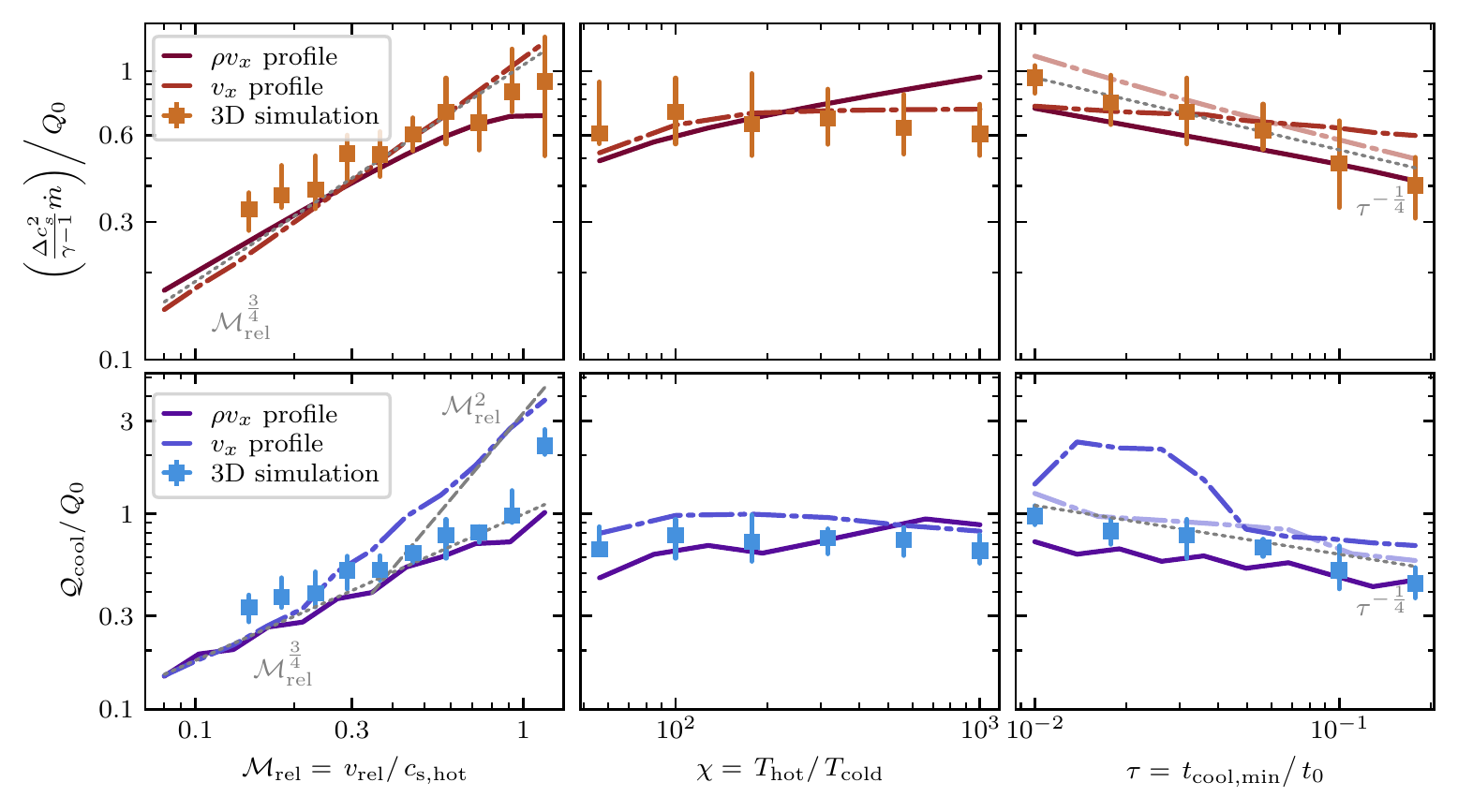}
\caption{Here we juxtapose the enthalpy flux and radiative cooling in mixing layers as a function of $\mathcal{M}_{\rm rel}$ (left column), $\chi$ (middle column), and $\tau$ (right column) from our analytic model with the cosine $v_x$ profile (dash-dotted lines), the cosine $\rho v_x$ profile (solid lines), and from the 3D simulations (scattered data points). To find the 3D simulation data points and error bars, we extract the time-dependent 3D simulation results between $t=2 t_{\rm mix}$ and $t=4 t_{\rm mix}$ (where $t_{\rm mix}=\left. \chi^{0.5} L \right/ v_{\rm turb}$, $v_{\rm turb}$ is measured in 3D simulations), which is when the thickness of the 3D TRML matches the thickness defined in our analytic model. The data points are the median data between these times, and the error bars denote the $2 \sigma$ temporal variations. For each point on the analytic model curves, we use values of $\PR$ and $f_{\nu}$ that best fits the corresponding 3D simulation using the scaling relationships shown in \autoref{fig:best fit Pr and f_nu for the vx and momentum profile}. The fiducial choices of parameters are $\mathcal{M}_{\rm rel}=0.75$, $\chi=10^2$, and $\tau=10^{-1.5}$. Across all six panels, the curves calculated from the two variations of the analytic model lie near the 3D simulation results and appears to bound reality. In the left column, the cosine $v_x$ profile is characterized by the upturn in radiative cooling, and the cosine $\rho v_x$ profile is characterized by the downturn in enthalpy flux. In the middle column, both the enthalpy flux and radiative cooling show weak dependence on $\chi$. In the right column, both the enthalpy flux and radiative cooling scales as $\tau^{\left. -1 \right/ 4}$, which is accurately reproduced by the cosine $\rho v_x$ profile and the $v_x$ profile with constant $\PR$ and $f_{\nu}$ (dash-dotted, semi-transparent lines).}
\label{fig:vx, momentum, 3D, energy sources and sinks vs. vrel}
\end{figure*}

Now that we have discussed how to find the best fit $\PR$ and $f_{\nu}$ values for our analytic model, we are ready to compare with 3D simulations. \autoref{fig:vx, momentum, 3D, energy sources and sinks vs. vrel} juxtaposes the enthalpy flux and radiative cooling computed in the two versions of our analytic model and in the 3D simulations as a function of $\mathcal{M}_{\rm rel}$, $\chi$ , and $\tau$. We emphasize that each point on the curves for the analytic models are computed using values of $\PR$ and $f_{\nu}$ that best fit the corresponding 3D simulation. The left column, where we vary $\mathcal{M}_{\rm rel}$, summarizes our analysis in \autoref{sec:vx profile, energy sources and sinks vs. vrel} and \autoref{sec:momentum profile, energy sources and sinks vs. vrel}. The cosine $v_x$ profile is characterized by the upturn in radiative cooling due to viscous heating near the cold phase, and the cosine $\rho v_x$ profile is characterized by the downturn in enthalpy flux due to viscous heating at intermediate temperature, and the 3D simulations lie somewhere in between, exhibiting both features. More specifically, at low $\mathcal{M}_{\rm rel}$ both the enthalpy flux and radiative cooling scales as $\mathcal{M}_{\rm rel}^{\frac{3}{4}}$, while at high $\mathcal{M}_{\rm rel}$ radiative cooling turns up to scale as $\mathcal{M}_{\rm rel}^2$ because the relative kinetic energy term is dominant. We see that, while results from both the $v_x$ and the $\rho v_x$ profiles are close to the 3D simulations, neither is a perfect match. We experimented with a few other choices (e.g., a cosine shear kinetic energy profile) but did not find any to be significantly better; for the rest of this paper, we will explore both profiles with the expectation that they bound reality.

Although we do not yet have a perfect analytic model that is able to encapsulate both features and match the 3D simulation perfectly, this investigation still gave us many insights and allowed us to build physical intuition. Most notably, we conclude that there are two means by which viscous dissipation can impact the mixing layer: (i) by heating gas at intermediate temperatures, which reduces the enthalpy flux from the hot phase as the shear Mach number increases while leaving the scaling between total cooling and the shear Mach number unchanged (see \autoref{sec:momentum profile, energy sources and sinks vs. vrel} and \autoref{fig:momentum profile, energy sources and sinks vs. vrel}), and (ii) by heating gas in the cold phase, which causes radiative cooling to follow a steeper scaling relationship with the shear Mach number as the shear Mach number approaches unity while leaving the scaling of enthalpy flux with the shear Mach number unchanged (see \autoref{sec:vx profile, energy sources and sinks vs. vrel} and \autoref{fig:vx profile, energy sources and sinks vs. vrel}). These two modes of viscous heating play a crucial role in setting the phase structure of the mixing layer and the mass, momentum, and energy transfer between the two phases.

Besides understanding how the enthalpy flux and radiative cooling in the mixing layers scale as $\mathcal{M}_{\rm rel}$, we also performed the same investigation while varying $\chi$ and $\tau$. The results are summarized in the middle and right columns of \autoref{fig:vx, momentum, 3D, energy sources and sinks vs. vrel}. The key point here is that results from the two variations of the analytic model bracket the 3D simulation data point, which support our earlier claim that the analytic models bound reality. More specifically, both the enthalpy flux and radiative cooling show very weak dependence with respect to $\chi$ and scales as $\tau^{\left. -1 \right/ 4}$. The $\tau^{\left. -1 \right/ 4}$ scaling is accurately reproduced by the cosine $\rho v_x$ profile prescription but not by the cosine $v_x$ profile prescription. We recognize this as a limitation of our cosine $v_x$ profile. However, in the two panels in the right column, we have also included two semi-transparent curves that are calculated using the cosine $v_x$ profile at constant (not best fit) values of $\PR$ and $f_{\nu}$ ($\PR=0.07$, $f_{\nu}=0.01$). The semi-transparent curves follow the $\tau^{\left. -1 \right/ 4}$ scaling perfectly.

\subsection{Matching the Shapes of the Phase Distributions with 3D Simulations}
\label{sec:matching phase distributions}

Being able to match the values of $\dot m$ and $\Qcool$ from 3D simulations tells us that we can successfully reproduce the mass transfer between the phases and the normalization of the cooling flux distribution. However, to be able to predict the detailed spectral features from the mixing layer, we usually need to rely on higher level features that depend on the shape of the cooling flux and column density distributions. In this section we explore the effectiveness of our analytic model in matching the shapes of the cooling flux and column density distributions from 3D simulations, which is a prerequisite for accurately predicting observational metrics.

In \autoref{sec:matching mdot and Qcool}, we have already described the procedure for choosing parameters in our analytic model to obtain the best fit to a 3D simulation. Now we simply compare the resulting cooling flux and column density distributions of our analytic model with that of the 3D simulation.

\begin{figure*}
\centering
\includegraphics[width=\textwidth]{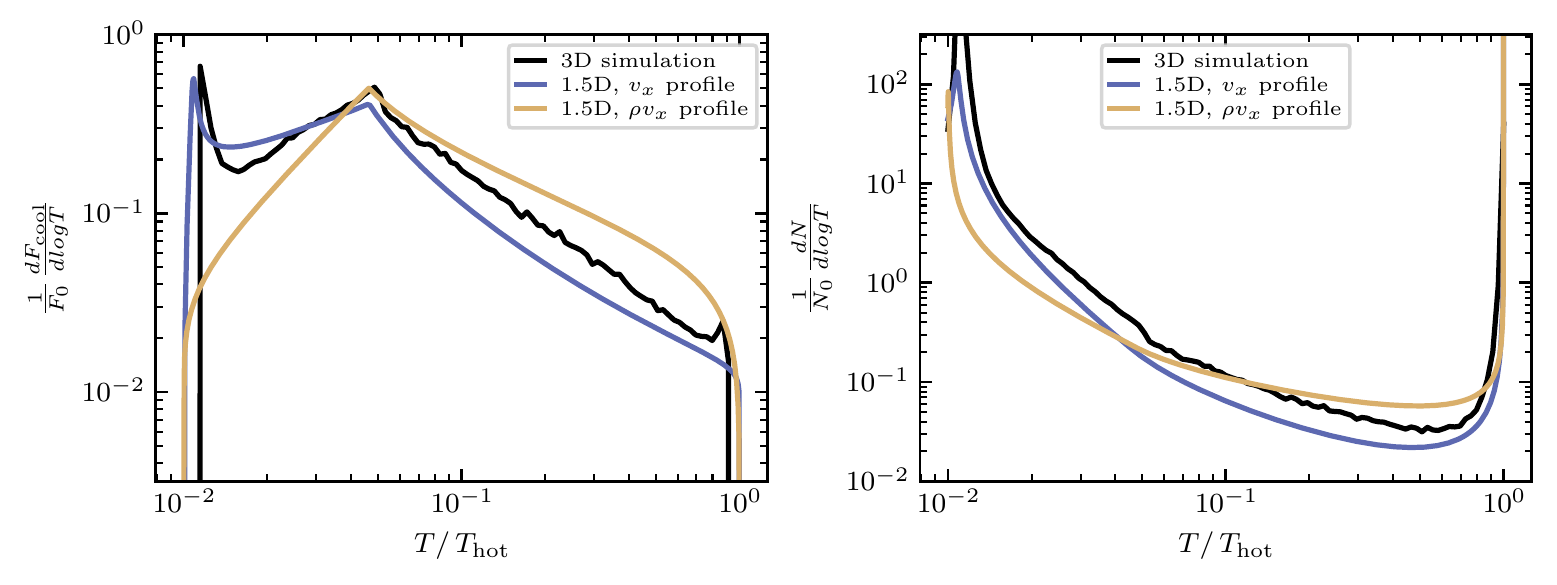}
\caption{Comparison of the cooling flux and column density distributions from the 3D simulation and the corresponding best fit analytic solution with the cosine $v_x$ and $\rho v_x$ profiles. The cooling flux distributions are normalized by $F_0 = \rho_{\rm hot} c_{\rm s,hot}^3$, and the column density distributions are normalized by $N_0 = \left. \rho_{\rm hot} L_0 \right/ m_H$. The 3D simulation has $\mathcal{M}_{\rm rel}=0.75$, $\chi=10^2$, $\tau=10^{-1.5}$, and its cooling flux and column density distributions are plotted in black. The analytic models have the same $\mathcal{M}_{\rm rel}$, $\chi$, and $\tau$, and the best fit $\PR$ and $f_{\nu}$ are obtained using the procedure described by \autoref{fig:schematic of best fit Pr and f_nu}. For the cosine $v_x$ profile, the best fit $\PR=0.79$ and best fit $f_{\nu}=0.013$. For the cosine $\rho v_x$ profile, the best fit $\PR=0.40$ and best fit $f_{\nu}=0.00073$. The phase distributions from the cosine $v_x$ profile (in blue) provides a better fit.}
\label{fig:matching phase distributions}
\end{figure*}

\autoref{fig:matching phase distributions} shows the cooling flux and column density distributions from a fiducial 3D simulation over-plotted with the best fit analytic models using both the $v_x$ and the $\rho v_x$ profile. The $v_x$ profile produces a close match except for a small mismatch in the slope of the cooling flux distribution at temperatures lower than the peak of the cooling curve and higher than the cold phase spike. The $\rho v_x$ profile, on the other hand, produces a less accurate match. This is expected because the $\rho v_x$ profile setup does not encapsulate the cold phase spike in the cooling flux distribution, which is an important feature that contributes significantly to the total amount of cooling $\Qcool$. To compensate for missing the cold phase spike, the cooling flux distribution produced by the $\rho v_x$ profile setup has a shallower slope at temperatures above the peak of the cooling curve, and as we saw in \autoref{sec:matching mdot and Qcool}, the resulting $\Qcool$ is remarkably accurate. Both profiles produce reasonable matches to the column density distribution, although the overall shape of the $v_x$ profile is a better match, for the reasons just mentioned.

\section{Discussion}
\label{sec:discussion}
\subsection{What have we learned about the physics of TRMLs}

Much recent research has provided essential insights to relevant problems using both 3D hydrodynamical simulations and analytic models. We now put our theory in the context of these existing works.

\cite{Gronke:2018} explained the importance of radiative cooling in TRMLs through cloud-crushing simulations (this had been seen phenomenologically before by \citealt{Marinacci:2010,Armillotta:2016}, and many times after, e.g., \citealt{Li:2020,Sparre:2020,Kanjilal:2021,Abruzzo:2021}). They presented a physical model to explain why sufficiently large clouds are able to grow in mass through radiative cooling of hot gas in the TRMLs. Similar results are also found in simulations of cold filaments embedded in a hot medium by \cite{Mandelker:2020b}.  \cite{Fielding:2020} demonstrated that the cooling is isobaric in fully resolved 3D simulations, leading them to argue that the hot gas inflow is driven by turbulence mixing rather than pressure gradients. In solutions produced by our analytic models, there is generally little or no pressure gradient, consistent with the argument of \cite{Fielding:2020}.

Additionally, \cite{Fielding:2020} found that the cold gas growth and entrainment is stronger under: (i) higher relative shear velocities ($\mathcal{M}_{\rm rel}$), (ii) higher density contrast ($\chi$), and (iii) more rapid cooling (smaller $\tau$). In the language of our analytic model, stronger cold gas growth is equivalent to a larger value of the mass flux $\dot m$. We see that $\dot m$ is positively correlated with $\mathcal{M}_{\rm rel}$ and $\chi$, and negatively correlated with $\tau$. These are consistent with the findings of \cite{Fielding:2020}.

\cite{Tan:2021} and \cite{Tan:2021b} explored a 1D model for TRML in parallel with their 3D simulation results. Their pioneering attempt of encapsulating the physics of complex TRMLs within a 1D model revealed a tremendous amount of insights. In particular, they experimented with a constant conductivity (with the constant value taken from Spitzer conductivity) and a temperature-dependent conductivity. Building from their work, the setup of our analytic model involves two main differences. First, we have included viscosity. Second, we use an effective turbulent conductivity and viscosity (as introduced in \autoref{sec: cond and turb}). These different choices reflects different understandings of the fundamental physical process at play. While \cite{Tan:2021} and \cite{Tan:2021b} believe that heat diffusion is dominated by thermal conduction, we believe that turbulence plays a more important role.

\cite{Yang:2022} also identified the importance of kinetic energy in the energy balancing argument. Their result is similar to the Bernoulli flux argument we introduced in \autoref{sec:Bernoulli flux}. However, a crucial difference is that \cite{Yang:2022} focused on the slowly cooling regime while we are primarily interested in the rapidly cooling regime. Additionally, \cite{Yang:2022} found a separation of turbulent and mixing zones in their TRMLs, which they argue to be a feature that develops in the high Mach number limit. In this work, we generally focus on Mach numbers that are much lower than what \cite{Yang:2022} explored, but we do observe a qualitatively similar separation of the turbulent and mixing zone in our cosine $v_x$ profile setup. However, the underlying physics is probably fundamentally different in the fast cooling regime, and we emphasize that the turbulent zone (or what we called "the viscous heating zone" in \autoref{sec:vx profile individual sol}) that we observe is in the cold phase, while the turbulent zone of \cite{Yang:2022} is in the hot phase.

We emphasize that our 1.5D analytic model has proven to be successful in matching the values of $\dot m$ and $\Qcool$ and the shape of the phase distributions from 3D simulations. This gives us confidence that the physical insights we gained from our analytic model also applies to the 3D simulations.

\subsection{A User's Guide to Our analytic Model} 
On the most practical level, the utility of our analytic model is that it allows users to obtain the key features of any TRML with any combination of parameters in a matter of seconds, without having to run the much more computationally expensive 3D simulations. For the convenience of the reader, we have put together several annotated notebooks that allows the reader to easily implement our analytic model. Following this \href{https://github.com/ziruichen11/1.5D_mixing}{URL}, the reader can find a series of four notebooks: two for calculating an analytic solution of a TRML for any given set of parameters (using both the cosine $v_x$ and $\rho v_x$ profile), and two for finding energy conserving pairs of $\PR$ and $f_{\nu}$ using the method outlined in \autoref{fig:schematic of best fit Pr and f_nu} (again, using both the cosine $v_x$ and $\rho v_x$ profile).

Finally, we emphasize here that besides being able to reproduce key features of the 3D simulations at a fraction of the computational cost, our analytic model is also valuable as a flexible, stand-alone model. Under our analytic model, it is easy to test how changing different parameters of our model affects the physics of the resulting mixing layer. For example, in \autoref{sec:phase distributions vs. beta_hi and lo}, we explore how changing $\betalo$ and $\betahi$ of the cooling curve (see \autoref{sec:cooling curve} for more details) affect the phase distributions of the mixing layer. Similar exercises on changing other parameters of the model can be easily performed as well.

\subsection{Links to Observations --- Calculating Ion Fractions and Column Densities Using Our Analytic Model}
\label{sec:links to observations, column density}

Due to its diffuse nature, the circumgalactic medium (CGM) is often mapped by absorption-line spectroscopy, where diffuse gas is detected through its absorption of light from bright background objects \citep[for a review see][]{Tumlinson:2017}. Absorption line observations of the CGM have, for example, revealed that the CGM is a dominant reservoir of baryons on galactic scales and can help explain the deficiency of baryons within the virial radius as inferred from the stellar mass to dark matter mass ratio \citep{Werk:2014}. Simultaneous measurements of multiple ions provide a valuable window into the turbulent and multiphase nature of the CGM across a broad range of redshifts \citep[e.g.][]{Werk:2016, Rudie:2019, Qu:2022}. Furthermore, measuring the column density of these tracer ions can reveal key characteristics of the galaxy formation process. For example, \cite{Tumlinson:2011} found that there is a correlation between the column density of \ion{O}{6} in the CGM and the specific star formation rate of galaxies, indicating that the CGM plays an important role in the evolution of galaxies. 

Absorption-line spectroscopy has been similarly powerful for understanding galactic winds \citep[for a review see][]{Veilleux:2020}. These observations have made it clear that galactic winds are a ubiquitous feature of star-forming galaxies \citep{Weiner:2009,Rubin:2014}, both in the local universe \citep[e.g.][]{Heckman:1990,Martin:1999}, and at high redshifts of $z\sim 2$ \citep[e.g.,][]{Rudie:2012}. These observations have been incredibly useful in determining how much mass is carried out of the galaxy by these winds and how this mass flux depends on galactic properties \citep[e.g.,][]{McQuinn:2019}. Galactic winds are known to be highly multiphase \citep{StricklandHeckman:2009}, which makes it essential to have a robust model for the phase structure in order to extract as much information as possible from these observations. 

In this Section, we discuss how our analytic model provides a quick and accurate way of calculating the ion fraction and column densities in a given TRML appropriate for either the CGM or a galactic wind. These calculations allow us to make predictions on the observable absorption features and better understand the composition and characteristics of the host galaxy.

Our analytic model provides us with the temperature, pressure, and density profiles of the mixing layer, which we can then use to calculate quantities that are directly tied to observations, including the ion fractions and ion column densities in the mixing layer.

To facilitate these calculations, we make two changes to our model. First, instead of the piece-wise power-law cooling curve introduced in \autoref{sec:cooling curve}, we adopt a realistic cooling curve. This reduces the flexibility of our model but allows for more accurate calculations. Second, we need to convert our results from a dimensionless form into physical units to ensure that our calculation is physically sensible. We have discussed how to do this conversion in \autoref{sec:unit conversion}.

For any given analytic solution to a TRML, we can proceed to calculate the ion fraction as a function of position in the mixing layer. We use the ion abundances table from {\tt TRIDENT} \citep{Hummels:2017}, which is tabulated using {\tt CLOUDY} \citep{CLOUDY}, under the assumption of ionization equilibrium assuming a background radiation field from \cite{Haardt:2012}. We note that this background radiation field is more applicable to the (outer) CGM and might need to be modified if one considers scenarios that are close to the disk or near quasars. The ion abundances are functions of redshift, metallicity, temperature, and density. Using the temperature and density profiles of a TRML obtained from our 1.5D model, one can calculate the ion fraction at any position in the mixing layer after assuming some redshift and metallicity values based on the physics of the scenario at hand. For the purposes of demonstration, we assume a redshift of 0 and solar metallicity in the remainder of the section and compute the ion fractions and column densities in a fiducial analytic solution of a TRML. We emphasize that this serves as a demonstration of the model capabilities rather than a definitive prediction. In fact, the solar metallicity assumption is more accurate for galactic winds and might need to be modified for the CGM.

Under these assumptions, we can calculate the ion fraction at any position in the mixing layer. Furthermore, the number density of an ion can be calculated by
\begin{align}
    n_{\rm ion} = n\times A_{\rm element} \times f_{\rm ion} ,
\end{align}
where $n$ is the total number density, $A_{\rm element}$ is the abundance of the atomic species of interest (assuming solar metallicity), and $f_{\rm ion}$ is the ion fraction. The column density of that ion can then be calculated by
\begin{align}
    N_{\rm ion} = \int n_{\rm ion} dz ,
\end{align}
where the integral runs through the entire mixing layer.

\begin{figure}
\centering
\includegraphics[width=\columnwidth]{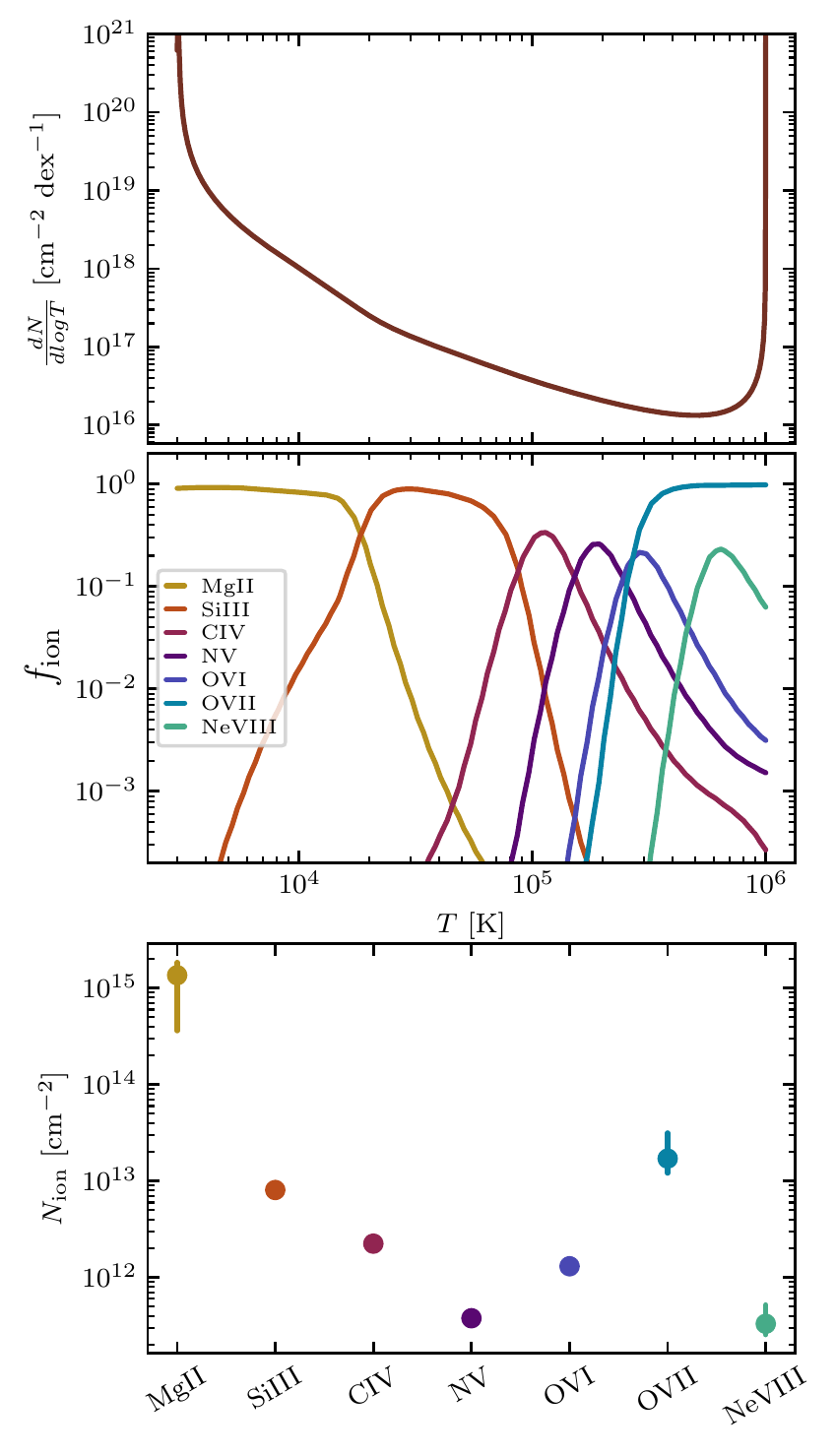}
\caption{Column density distribution from an analytic solution (top), the corresponding ion fraction as a function of temperature for a selected list of ions (middle, assuming a redshift of 0 and solar metallicity), and column density of these ions in the mixing layer (bottom). The solution uses a cosine $v_x$ profile and a realistic cooling curve. The system has $p_{\rm hot}/ k_B = 1000 {\rm K \, cm}^{-3}$, $T_{\rm hot}=10^6 {\rm K}$, $\mathcal{M}_{\rm rel}=0.75$, $\PR=0.011$, $f_{\nu}=10^{-2}$, and $\tau=10^{-1}$. The cold phase temperature is set by thermal equilibrium with the UVB, which corresponds to $\chi=332$. With the solution for the temperature, density, and pressure profiles for the mixing layer we calculate the ion fraction and column density of several commonly observed ions. The data points in the bottom panel are calculated assuming $\delta T = 0.01$, and the error bars represent values at $\delta T=0$ and $\delta T = 0.1$. (see \autoref{sec:links to observations, column density} for more details) This Figure can be reproduced by the reader with any choice of parameters and ions of interest using our annotated notebook that is made publicly available at this \href{https://github.com/ziruichen11/1.5D_mixing/tree/main/column_density_calculation}{URL}}
\label{fig:mass distribution and column density}
\end{figure}

In \autoref{fig:mass distribution and column density}, we plot the column density distribution generated by our analytic model, the ion fractions, and the column densities of a selected list of ions. We note that the column density of the ions that probe either the cold phase (e.g. \ion{Mg}{2}) or the hot phase (e.g. \ion{O}{7}) is dependent on the amount of cold or hot gas that the TRML connects. Physically, this corresponds to probing the interior of the cold cloud or the extent of the volume-filling hot gas. Our analytic model is for the TRML at intermediate temperatures only and does not provide a robust prediction for the cold and hot gas content on either side of the TRML. This can also be seen from the fact that \autoref{eq:dN_dlogT} for the column density distribution $\left. {\rm dN} \right/ {\rm dlogT}$ diverges at the cold and hot end of the solution. This is a limitation, but not an inaccuracy, of our model. To understand how this limitation affects our column density calculations, we calculate the column density of a given ion between $T = (1 + \delta T) T_{\rm cold}$ and $T = (1 - \delta T) T_{\rm hot}$, where $\delta T$ is a small, positive constant. In the bottom panel of \autoref{fig:mass distribution and column density}, we plot the column densities calculated using $\delta T = 0.01$ and use the error bars to indicate column density values at $\delta T = 0$ (upper bound) and $\delta T = 0.1$ (lower bound). As shown by the error bars, the ions that are the most abundant at either the cold or the hot phase (e.g. \ion{Mg}{2}, \ion{O}{7}) are mildly affected by the choice of $\delta T$, while most of the other ions are unaffected.

We note that the process described above assumes ionization equilibrium, which may be significantly in error in some circumstances \citep{2013MNRAS.434.1043O}. An additional point of caution in interpreting \autoref{fig:mass distribution and column density} is an inconsistency in accounting for self-shielding in our calculations. Physically, as the ionizing photons make their way in from the hot phase towards the center of the cold cloud they will be absorbed. Eventually the photons will be fully absorbed and the assumption of photo-ionization equilibrium will no longer be valid. Where self-shielding becomes an important factor will depend on the details of the mixing layer profile, but will primarily impact only the coldest gas and therefore the ions with the lowest ionization potential. This effect is ignored in our tabulated cooling function, but is approximately taken into account in the \citet{Ploeckinger:2020} tables we adopt. It will be straightforward in the future to self-consistently account for this by using {\tt CLOUDY} to calculate the equilibrium ionization state for the specific density and temperature profiles from our TRML calculations.

We have prepared an annotated notebook that produces figures similar to \autoref{fig:mass distribution and column density} with any ion of interest. The notebook can be found at this \href{https://github.com/ziruichen11/1.5D_mixing/tree/main/column_density_calculation}{URL}. We emphasize that these column densities are for a specific choice of pressure, velocity, metallicity, and radiative background and so serves as a demonstration of the model capabilities rather than a definitive prediction.

\subsection{Links to Observations --- Calculating Surface Brightness Using Our Analytic Model}
\label{sec:links to observations, surface brightness}

Although harder to come by than absorption line observations, direct observations of the emission from galactic winds and the CGM provides an extremely powerful probe. Absorption lines generally give us a single pencil beam view per system, but emission observations have the potential to provide valuable insight into the spatial structure of these systems. New instruments are making emission observations much more readily accessible. Specifically, KCWI and MUSE are providing a promising path forward to greatly improve our understanding galactic winds and the CGM as has been recently demonstrated \citep[e.g.,][]{Hayes:2016, Rupke:2019, Burchett:2021, ReichardtChu:2022}. As these new observations become available our need for robust models to interpret them becomes paramount.

As discussed in \autoref{sec:links to observations, column density}, our analytic solution provides us with the temperature, pressure, and density profiles of a mixing layer, which, after being converted to physical units (see \autoref{sec:unit conversion} for more details), allows us to determine the emissivity ($e$), and thus the surface brightness of any given emission line of interest. To find the emissivity of a given emission line, we use the properties of shielded gas tabulated by \cite{Ploeckinger:2020}, which uses {\tt CLOUDY} \citep{CLOUDY} to calculate the equilibrium cooling and heating rates in the presence of a UV background based on \cite{Faucher-Giguere:2020}, interstellar radiation and cosmic rays that depend on local gas properties
through the Kennicutt-Schmidt (KS) relation \citep{Kennicutt:1998}, and dust. The emissivity tables in \cite{Ploeckinger:2020} are given as a function of redshift, temperature, metallicity, and density. We assume a redshift of 0 and solar metallicity (see \autoref{sec:links to observations, column density} for a discussion of these choices), which means the emissivities are functions of temperature and density only ($e(T,n)$). Given the temperature and density profiles from our analytic solution, we can determine the emissivity of any given emission line as a function of position in the mixing layer. The surface brightness of that emission line can be calculated by
\begin{align}
    {\rm SB} = \int e{\left(T(z),n(z)\right)} dz ,
\end{align}
where SB is the surface brightness, $e$ is the emissivity, and the integral runs through the entire mixing layer. 

\begin{figure}
\centering
\includegraphics[width=\columnwidth]{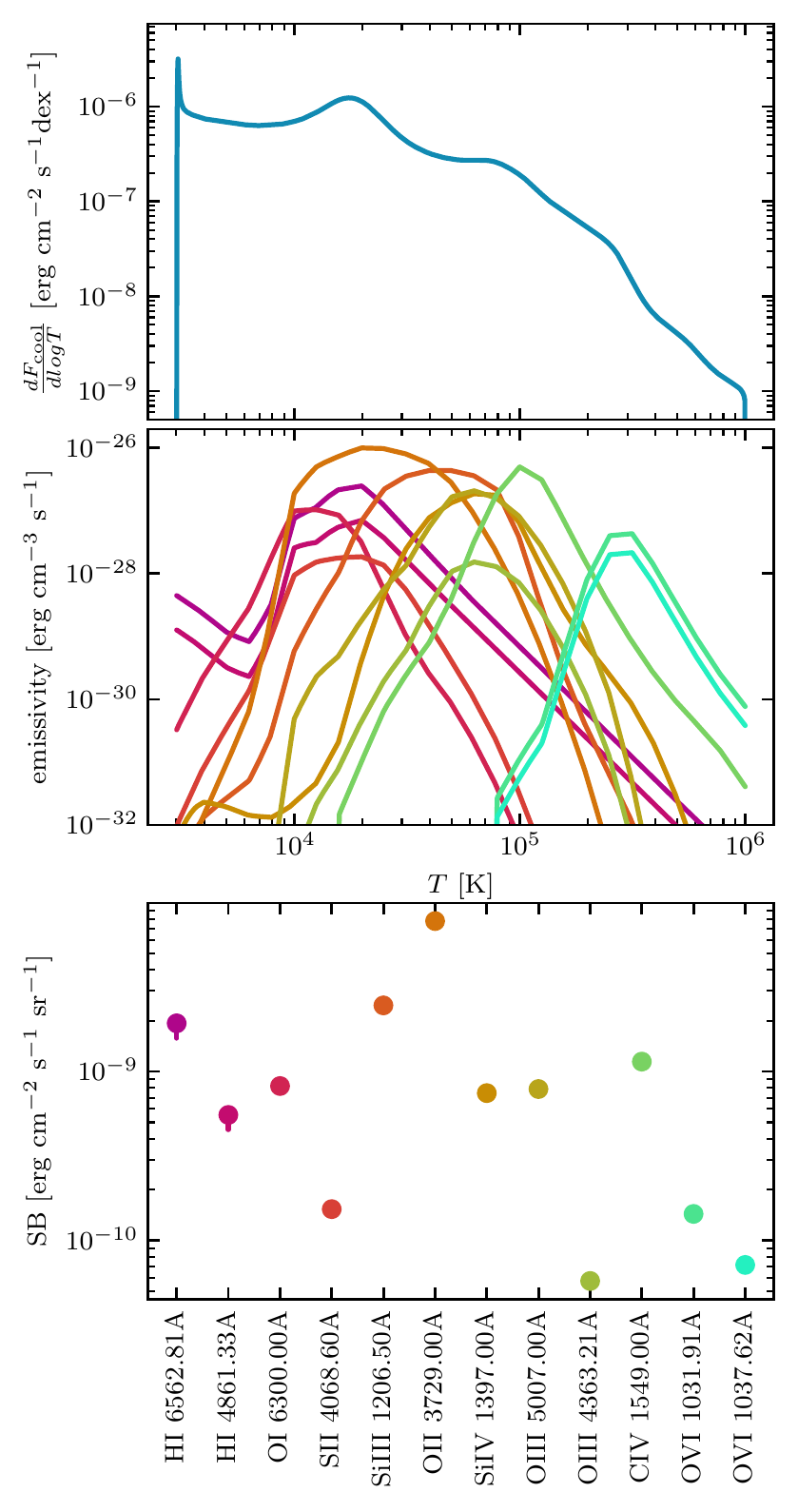}
\caption{Cooling flux distribution from an analytic solution (top), the corresponding emissivity as a function of temperature for a selected list of emission lines (middle, assuming a redshift of 0 and solar metallicity), and surface brightness of these emission lines in the mixing layer (bottom). The analytic solution we use here is the same as what we described in \autoref{fig:mass distribution and column density}. The temperature and density profiles from the analytic solution allows us to calculate the emissivity and surface brightness of any emission line, as discussed in \autoref{sec:links to observations, surface brightness}. The data points in the bottom panel are calculated assuming $\delta T = 0.01$, and the error bars represent values at $\delta T=0$ and $\delta T = 0.1$. (see \autoref{sec:links to observations, surface brightness} for more details) This Figure can be reproduced by the reader with any choice of parameters and ions of interest using our annotated notebook that is made publicly available at this \href{https://github.com/ziruichen11/1.5D_mixing/tree/main/surface_brightness_calculation}{URL}}
\label{fig:cooling distribution and surface brightness}
\end{figure}

In \autoref{fig:cooling distribution and surface brightness}, we plot the cooling flux distribution generated by our analytic model, the emissivities, and the surface brightness of a selected list of emission lines. As in \autoref{sec:links to observations, column density}, we calculate the surface brightness of a given ion between $T = (1 + \delta T) T_{\rm cold}$ and $T = (1 - \delta T) T_{\rm hot}$, where $\delta T$ is a small, positive constant. In the bottom panel of \autoref{fig:cooling distribution and surface brightness}, we plot the column densities calculated using $\delta T = 0.01$ and use the error bars to indicate column density values at $\delta T = 0$ (upper bound) and $\delta T = 0.1$ (lower bound). As shown by the error bars, most of the emission lines that are being plotted are unaffected by the choice of $\delta T$.

We have also prepared an annotated notebook that produces similar figures with any emission of interest. The notebook can be found at this \href{https://github.com/ziruichen11/1.5D_mixing/tree/main/surface_brightness_calculation}{URL}.

\subsection{Caveats section}
In this section, we discuss some physical effects that are not fully captured in our model and identify possible directions of future work.

\emph{Realistic Thermal Conduction and Viscosity}: In our analytic model we rely solely on the effective turbulent thermal conduction and viscosity, however in reality there will also be real conduction and viscosity that arises from particle collisions \citep{Spitzer:1962}. The strength and scaling of the conduction and viscosity coefficients in the tenuous plasmas that we are considering is uncertain. There can be large suppression possible due to the anisotropic motion of charged particles along the magnetic fields and by the excitation of resonant whistler waves \citep{Roberg-Clark:2016, Komarov:2018, Drake:2021}. Nevertheless, the impact of these missing process can be assessed and could be added in the future in a straightforward manner. The standard Spitzer conductivity of, for example, a system with $p_{\rm hot}/ k_B = 1000 {\rm K \, cm}^{-3}$ is $\kappa_{\rm Sp} (10^6 {\rm K}) = 10^{24.5}$ cm$^{-1}$ s$^{-1}$ and $\kappa_{\rm Sp} (10^4 {\rm K}) = 10^{19.5}$ cm$^{-1}$ s$^{-1}$. We can directly compare this to the effective turbulent conductivity of this system if $\vrel = 100$ km/s, and $h = 100$ pc, which yields a characteristic value of $\kappa_{\rm turb} \approx n_{\rm mix} h \vrel \approx 10^{25.5}$ cm$^{-1}$ s$^{-1}$, where $n_{\rm mix} = (p_{\rm hot}/ k_B) / T_{\rm mix} = (p_{\rm hot}/ k_B) / \sqrt{T_{\rm hot} T_{\rm cold}}$. Therefore, we expect that the effective turbulent conductivity will dominate even over full strength Spitzer conduction. This is also true for the viscosity since the Spitzer Prandtl number is $\PR_{\rm Sp} = \sqrt{m_e / m_p} \approx 1/42$, while the effective turbulent Prandtl number is $\PR \approx 1/10$. Nevertheless, there may be appreciable changes to the detailed structure of a TRML when realistic conduction and viscosity are included, and it would be straightforward to add to our existing TRML model.

\emph{Fractal Nature of TRMLs}: The fractal nature of the thin corrugated surface where all the cooling takes place in 3D simulations of TRMLs identified by \cite{Fielding:2020} is not directly captured in our analytic model. Remarkably, missing this physics does not prevent the analytic model from reproducing the mass flux and phase distributions of the 3D simulations. This suggests that our effective turbulent conductivity and viscosity prescription and the introduction of the ``model-specific" parameters $\PR$ and $f_{\nu}$ can be easily tuned to produce the same cooling volume.

\emph{Non-equilibrium Cooling}: The cooling curve we are using assumes equilibrium cooling, which can be incorrect when cooling is rapid \citep[e.g.,][]{2017ApJS..228...11G}. However, our analytic model is flexible enough that it should be relatively straightforward to implement a setup with non-equilibrium cooling.

\emph{Magnetic Fields and Cosmic Rays}: Our analytic model does not include the effects of magnetic fields and cosmic rays. These components can suppress Kelvin-Helmholtz instabilities \citep{Ji:2019} and possibly affect the phase structure and mass transfer between the phases.

\section{Summary}
In this work, we present an analytic, steady-state, 1.5D model for Turbulent Radiative Mixing Layers (TRMLs) that includes, for the first time, a physically motivated model of turbulent conduction and viscosity that is key for setting the flow of gas through the TRML. We first summarize our trajectory in this paper before reviewing the key results.

In \autoref{sec:methods}, we start from mass, momentum, and energy conservation to derive a system of second order, coupled differential equations that describe the TRML. We then define the dimensionless numbers, $\chi$, $\mathcal{M}_{\rm rel}$, and $\tau$, that most strongly affect the structure of TRMLs and discuss conversion between code and physical units. Finally, we introduce other features of our model, including a piece-wise power-law cooling curve and, most crucially, an effective turbulent conductivity and viscosity that are proportional to the shear velocity gradient ($\left. d v_x \right/ dz$). We explore two forms for the shear velocity, one motivated by the $v_x$ profile in simulations, and the other by the $\rho v_x$ profile. The introduction of a physical model for the turbulent conduction and viscosity which depends on a prescribed velocity model is a key new element of the model and allows for a much more self-consistent solution.

We solve this system of equations numerically to obtain solutions, and in \autoref{sec:results}, we present those results. We first examine individual solutions to the TRML obtained from the $v_x$ profile and identify the viscous heating zone that takes up a tiny fraction of the temperature space near the cold phase but a significant fraction of the position space. Next, we analyze the physics of TRMLs through the integrated energy sources and sinks and argue that the criterion $\Hvisc= \left. \dot m v_{\rm rel}^2 \right/ 2$ must be satisfied for energy conservation. 

By looking at how the total cooling in the mixing layer ($\Qcool$) and the enthalpy flux from the hot phase vary as a function of $\mathcal{M}_{\rm rel}$, we conclude that there are two components of viscous dissipation in the TRML that controls the phase structure and mass transfer between the phases: viscous dissipation at intermediate temperatures reduces the enthalpy flux from the hot phase, and viscous dissipation at the cold phase enhances radiative cooling. The energy conservation criterion $\Hvisc= \left. \dot m v_{\rm rel}^2 \right/ 2$ reduces one degree of freedom in the parameter space of our analytic model. In other words, $\PR$ and $f_{\nu}$ \footnote{We remind the readers here that $\PR$ is the ratio of the viscosity $\mu$ and the conductivity $\kappa$, and $f_{\nu}$ is the constant prefactor in our definition of $\kappa$ that serves to define an effective transverse lengthscale.}, the two parameters that are introduced by our analytic model and not present in the 3D simulation, can effectively be reduced to one. Comparing with 3D simulation results allows us to identify the pair of $\PR$ and $f_{\nu}$ that best matches the $\dot m$ and $\Qcool$ for each of the two forms of the shear velocity profile (cosine fits to $v_x$ and  $\rho v_x$). (see \autoref{fig:vx profile, match mdot and Q_cool} and \autoref{fig:momentum profile, match mdot and Q_cool} for more details). The best fit analytic solutions also produces a remarkably good fit to the phase distributions, especially with the cosine $v_x$ profile setup. We show this result in \autoref{fig:matching phase distributions}. In general, we expect the best fit analytic solutions produced by the two forms of the shear velocity profile to bound reality.

In \autoref{sec:discussion}, we put this work in the context of the existing literature. We also link our analytic model with observations and discuss the calculation of ion fractions and ion column densities in the TRML. For the convenience of the reader, we have uploaded annotated notebooks for implementing our analytic model and calculating column densities of ions.

\vspace{3mm}\noindent Our key findings are as follows:
\begin{itemize}
    \item Our model provides essential insights into the physics that controls the evolution of TRMLs. Specifically, any TRML that satisfies energy conservation must obey $\Hvisc= \left. \dot m v_{\rm rel}^2 \right/ 2$, where $\Hvisc$ is the amount of viscous heating in the TRML, $\dot m$ is the mass flux between the hot and cold phases, and $v_{\rm rel}$ is the relative shear velocity between the phases. This viscous dissipation of relative kinetic energy is particular important in balancing radiative cooling as the shear Mach number approaches unity.
    \item By introducing two ``model-specific" parameters, $\PR$ and $f_{\nu}$, and calibrating their values using results from 3D simulations, we are able to reproduce the mass flux $\dot m$, total cooling $\Qcool$, and phase structure of the computationally expensive 3D simulations \citep{Fielding:2020}. 
    \item We outline a quick and accurate way of calculating the column density of ions and surface brightness of emission lines in a given TRML using our model. The flexibility of our model allows for such calculations to be carried out for a wide range of physical conditions (pressure, metallicity, shear velocity), and we provide a demonstration of such calculations using a fiducial set of parameters.
\end{itemize}

\acknowledgements
We thank Peng Oh, Max Gronke, Matthew Abruzzo, Brent Tan, and Eve Ostriker for useful discussions. ZC thanks the Department of Astronomy and Department of Physics at Columbia University for four years of undergraduate education and the Simons Foundation for its hospitality. GLB acknowledges support from the NSF (AST-2108470, XSEDE grant MCA06N030), NASA TCAN award 80NSSC21K1053, and the Simons Foundation (grant 822237). DBF is supported by the Simons Foundation through the Flatiron Institute. This work is supported by the Simons Collaboration on “Learning the Universe”.

\appendix

\section{Details of Our Cooling Curve}
\label{sec:cooling curve details}

As mentioned in \autoref{sec:cooling curve}, we use a piece-wise power-law cooling curve that is intended to reproduce the overall shape of the realistic cooling curves. Additionally, we include a heating term such that the cooling curve has two equilibria at both the cold and the hot phase temperatures. This heating term improves numerical performance but has no appreciable impact on our results.

Our cooling curve is given by
\begin{ceqn}
\begin{align}
    \Edotcool = \frac{1}{\gamma-1} &\frac{P_{\rm hot}}{t_{\rm cool, min}} \lr{\frac{P}{P_{\rm hot}}}^2 \biggl(\lr{\frac{T}{\Tpeak}}^{-\beta} \nonumber\\
    &- c_{\rm heat} \lr{\frac{T}{\Tpeak}}^{\alpha_{\rm heat}}\biggr) ,
    \label{eq:edot_cool}
\end{align}
\begin{align}
    {\rm where} \quad &\beta = \begin{cases}
    \betalo & \text{if } T \leq \Tpeak \\
    \betahi & \text{if } T > \Tpeak  \nonumber
\end{cases} ,
\end{align}
\begin{align}
    & c_{\rm heat} = \lr{\frac{T_{\rm cold}}{\Tpeak}}^{\left(\betahi - \betalo \right)\left(1+ \frac{\log_{10} \left(\left. T_{\rm cold} \right/ T_{\rm peak} \right)}{\log_{10} \chi}  \right)} \nonumber , \\
    {\rm and} \quad & \alpha_{\rm heat} =\left(\betalo - \betahi \right)\left(\frac{\log_{10} \left(\left. T_{\rm cold} \right/ T_{\rm peak} \right)}{\log_{10} \chi}  \right) - \betahi \nonumber
\end{align}
\end{ceqn}

Note that $\tau$ is the dimensionless number that controls the normalization of the cooling rate, which we introduced in \autoref{sec:dimensionless numbers}, and $\Tpeak$ is the temperature where the cooling rate peaks and where the cooling time, defined to be
\begin{ceqn}
\begin{align}
    \tcool = \frac{\left. P \right/ \lr{\gamma-1}}{\Edotcool},
    \label{eq:cooling time}
\end{align}
\end{ceqn}

is the shortest.

As shown in \autoref{eq:edot_cool} and \autoref{eq:cooling time}, the dimensionless number $\tau$ serves as a normalization to both the cooling rate and the cooling time. In 3D simulations, this normalization is set by the shear time $t_{\rm sh} = \left. L \right/ v_{\rm rel}$, where $L$ is the shear layer length, which is well-defined in 3D simulations. However, here in our 1.5D model we do not have a transverse lengthscale, which precludes us from implementing the same setup. Instead, we use the characteristic lengthscale introduced in \autoref{eq: lengthscale} to define the dimensionless number $\tau$, which serves as the normalization in our setup.

\section{Details of the Numerical Integration}
\label{sec:integration details}
We seek to numerically integrate 2 coupled differential equations to solve for the z-velocity ($v_z$) and temperature profiles. The equations at hand are second order differential equations in $v_z$ and $T$ (\autoref{eq:z equation} and \autoref{eq:T equation}) that we obtained from the steady state fluid equations. Our general strategy is to use scipy's solve-ivp integrator to perform the numerical integration after defining appropriate initial conditions. 

We are primarily interested in condensation solutions that flow from the hot to the cold phase. The natural choice of the direction of integration is to follow the direction of gas flow: from hot to cold. This means we can deal with positive values of $v_z$ and the mass flux $\dot m$ and avoid potential numerical difficulties.

Now we are left with defining the initial conditions, which consists of $T$ and $v_z$ at the hot phase, and the gradients of $T$ and $v_z$ at the hot phase. The hot phase temperature and pressure is set to 1 for convenience, and the hot phase density can be calculated through $\rho_{\rm hot} = \left. P_{\rm hot} \right/ T_{\rm hot}$. Then for a given choice of the mass flux $\dot m$, the hot phase $v_z$ can be calculated through $v_{\rm z,hot} = \left. \dot m \right/ \rho_{\rm hot}$. As for the initial gradients, we note that by definition our cooling curve has an equilibrium at the hot phase, so to produce non-trivial solutions that actually depart from the hot phase, we need to give the flow an initial "nudge" to push it away from the equilibrium and evolve towards the cold phase. To do so, we define the initial $\left. dT \right/ dz$ to be a small negative number. The choice of the initial gradients does not affect the solutions we obtain as long as the initial gradients are non-zero and reasonably small (approximately $10^{-6}$).

With the initial conditions defined, we proceed to numerically integrate our system of differential equations for a given value of mass flux $\dot m$. We use scipy's solve-ivp integrator with an rtol of $3 \times 10^{-14}$ and atol of $10^{-11}$ and terminate the integration if $T$ drops below the cold phase $T$ calculated through the density contrast $\chi$ or if $T$ exceeds the hot phase $T$ (more on why this can happen in the following paragraphs). 

Note that in the above description, the mass flux $\dot m$ seems to be a free parameter. Now we impose the additional constraint that the temperature profile needs to be flat ($\left. dT \right/ dz = 0$) at both the hot and cold phases by definition. For any given choice of parameters, there's only one value of the mass flux $\dot m$ that allows for that. We call that the eigenvalue of $\dot m$ (${\dot m}_{\rm eigen}$). 

To find the eigenvalue of $\dot m$, it is crucial to realize that when $\dot m$ is greater than ${\dot m}_{\rm eigen}$, the $T$ profile shoots down to the cold phase with a steep slope, meaning that $\left. dT \right/ dz < 0$ at the end of the solution; when $\dot m$ is less than ${\dot m}_{\rm eigen}$, the $T$ profile barely reaches the cold phase before bouncing back up, and since we terminate the solution when $T$ exceeds the hot phase $T$, we end up with $\left. dT \right/ dz > 0$ at the end of the solution in this case. Given the different signs of $\left. dT \right/ dz$ at the end of the solution in these two cases and the fact that we want to find a solution with $\left. dT \right/ dz = 0$ at the cold phase, we adopt a bisection method in the parameter space of $\dot m$ to pinpoint ${\dot m}_{\rm eigen}$. (see \autoref{fig:T profiles at different mdot} for more details.) Specifically, we start off from the maximum possible value of $\dot m$ given by ${\dot m}_{\rm crit} = \left. P_{\rm hot} \right/ \sqrt{T_{\rm hot}} = \rho_{\rm hot}  \sqrt{T_{\rm hot}}$ and record the sign of $\left. dT \right/ dz$ at the end of the solution with $\dot m = {\dot m}_{\rm crit}$. Then we decrease our guess of $\dot m$ by 10\% and record the sign of $\left. dT \right/ dz$ at the end of the new solution. We continue this process until we notice a sign change. Then ${\dot m}_{\rm eigen}$ must be sandwiched between the two most recent guesses for $\dot m$. Finally, we use scipy's optimize.root-scalar with rtol=$10^{-14}$ to find the $\dot m$ value that corresponds to the solution with $\left. dT \right/ dz = 0$ at the cold phase. This is the eigenvalue of $\dot m$ we are looking for.

\section{Phase Distribution as a Function of $\betalo$ and $\betahi$}
\label{sec:phase distributions vs. beta_hi and lo}

Besides serving as a computational efficient counterpart to the 3D simulation, our analytic model also has value as a flexible stand-alone framework. In particular, we can easily test how changing different parameters of our model affects the structure of the phase distributions. In this section, we focus on understanding how $\betalo$ and $\betahi$, parameters of the cooling curve (see \autoref{sec:cooling curve} for more details), affect the shape of the phase distributions.

\begin{figure*}
\centering
\includegraphics[width=\textwidth]{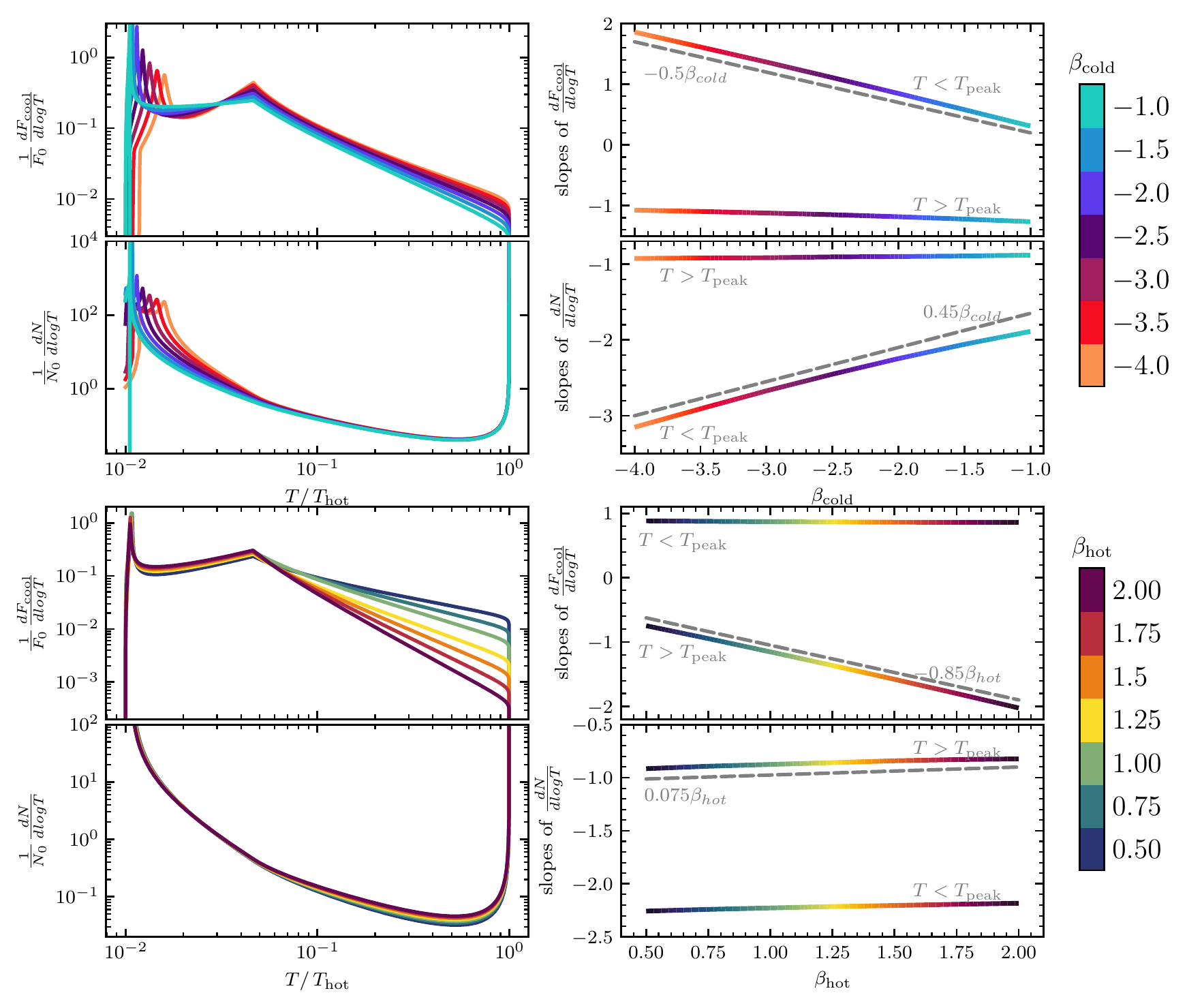}
\caption{In the top half of the figure we show the cooling flux distributions, column density distributions, and the power-law slope of these phase distributions for a series of mixing layers with different $\betalo$. We use the cosine $v_x$ profile setup here, and all other parameters are held constant: $\mathcal{M}_{\rm rel}=0.75$, $\tau=10^{-1}$, $\chi=10^2$, $\PR=0.07$, $f_{\nu}=10^{-2}$, and $\betahi=1$. The cooling flux distributions are normalized by $F_0 = \rho_{\rm hot} c_{\rm s,hot}^3$, and the column density distributions are normalized by $N_0 = \left. \rho_{\rm hot} L_0 \right/ m_H$. Note that $\PR$ and $f_{\nu}$ are selected such that energy conservation is satisfied. The color scheme denotes the different choices of $\betalo$. Generally speaking, changing $\betalo$ affects the power-law slopes of the cooling flux and column density distribution at $T < \Tpeak$ linearly and has almost no effect on the power-law slopes at $T > \Tpeak$. The bottom half of the figure is identical to the top half other than that we fix $\betalo=-2$ and vary $\betahi$. Varying $\betahi$ affects the power-law slopes of the cooling flux and column density distributions at $T > \Tpeak$ linearly and has almost no effect on the power-law slopes at $T < \Tpeak$.}
\label{fig:vx profile, phase distributions vs. beta_cold and beta_hot}
\end{figure*}

\autoref{fig:vx profile, phase distributions vs. beta_cold and beta_hot} is made using the cosine $v_x$ profile setup and shows how the phase distributions vary as a function of $\betalo$ and $\betahi$. Note that the values of $\PR$ and $f_{\nu}$ are carefully selected such that all the mixing layers being plotted obey energy conservation.

As we vary $\betalo$, the slopes of the phase distributions at $T < T_{peak}$ changes linearly with it, while the slopes at $T > T_{peak}$ remain unchanged. The opposite holds true when we vary $\betahi$, although the slope of the column density distribution at $T > T_{peak}$ seems to change only very slightly.

\begin{figure*}
\centering
\includegraphics[width=\textwidth]{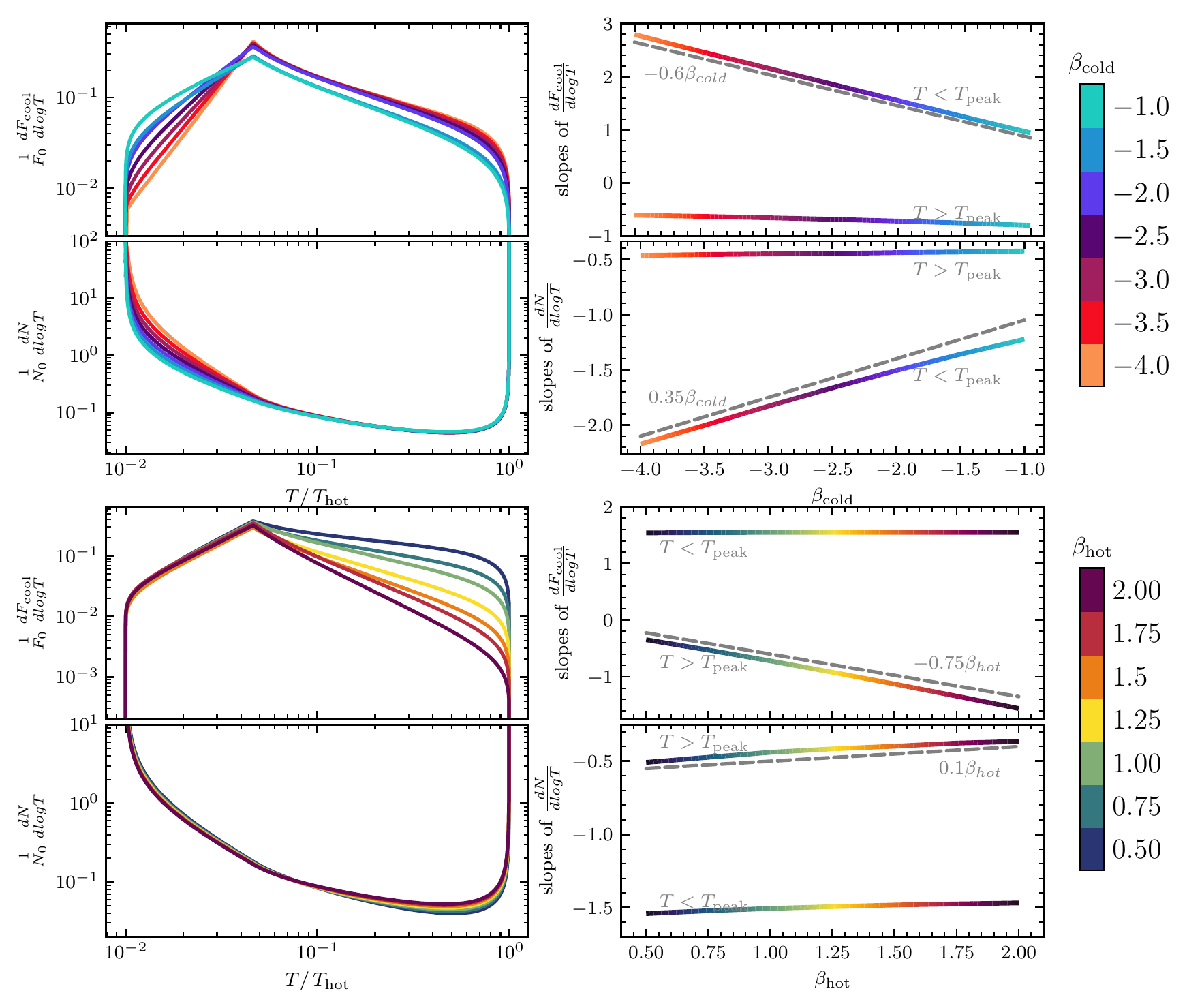}
\caption{Same as \autoref{fig:vx profile, phase distributions vs. beta_cold and beta_hot}, but with the cosine $\rho v_x$ profile.}
\label{fig:momentum profile, phase distributions vs. beta_cold and beta_hot}
\end{figure*}

\autoref{fig:momentum profile, phase distributions vs. beta_cold and beta_hot} shows the same experiment but with the cosine $\rho v_x$ profile setup. The same conclusions hold. 

\bibliography{references}

\end{document}